\def\R{{\bf R}}
\def\r{{\bf r}}
\def\tento#1{\times 10^{#1}}
\def\n{n}
\def\m{m}
\def\imp{\mapsto}
\def\reverseimp{\leftarrow}
\def\och{\&}
\def\eller{\vee}
\def\icke{\sim}
\def\sheffer{|}

\def\ie{{\frenchspacing i.e.}}
\def\eg{{\frenchspacing e.g.}}
\def\etc{{\frenchspacing etc.}}

\def\ed{\end{document}}

\def\rf#1;#2;#3;#4;#5 {\par#1, {\it #3} {\bf #4}, #5 (#2). \par}

\def\beq#1{\begin{equation}\label{#1}}
\def\eeq{\end{equation}}
\def\beqa#1{\begin{eqnarray}\label{#1}}
\def\eeqa{\end{eqnarray}}
\def\eq#1{equation~(\ref{#1})}
\def\Eq#1{Equation~(\ref{#1})}

\def\fig#1{Figure~\ref{#1}}
\def\Fig#1{Figure~\ref{#1}}

\def\spose#1{\hbox to 0pt{#1\hss}}
\def\simlt{\mathrel{\spose{\lower 3pt\hbox{$\mathchar"218$}}
     \raise 2.0pt\hbox{$\mathchar"13C$}}}
\def\simgt{\mathrel{\spose{\lower 3pt\hbox{$\mathchar"218$}}
     \raise 2.0pt\hbox{$\mathchar"13E$}}}
\def\simpropto{\mathrel{\spose{\lower 3pt\hbox{$\mathchar"218$}}
     \raise 2.0pt\hbox{$\propto$}}}

\documentstyle[pra,twocolumn,aps,epsf,rotate]{revtex}
\begin{document}

\preprint{}

\title{Is ``the theory of everything" merely the 
ultimate ensemble theory?}

\author{Max Tegmark}

\address{Institute for Advanced Study, Olden Lane,
Princeton, NJ 08540; max@ias.edu
}

\maketitle

\begin{abstract}              
We discuss some physical consequences
of what might be called ``the ultimate ensemble theory", where not only 
worlds corresponding to say different sets of initial data 
or different physical constants are 
considered equally real, but also worlds ruled by altogether different 
equations.
The only postulate in this theory is that 
all structures that exist mathematically exist 
also physically, by which we mean that 
in those complex enough to contain 
self-aware substructures (SASs),
these SASs will subjectively perceive themselves as existing in a 
physically ``real" world. 
We find that it is far from clear that this simple theory, which has
no free parameters whatsoever, is observationally ruled out. 
The predictions of the theory take the
form of probability distributions for the outcome of experiments,
which makes it testable. In addition, it may be possible to rule
it out by comparing its {\it a priori} predictions for 
the observable attributes 
of nature (the particle masses, the dimensionality of spacetime, \etc)
with what is observed. 
\end{abstract}


\makeatletter
\global\@specialpagefalse
\def\@oddfoot{
\ifnum\c@page>1
  \reset@font\rm\hfill \thepage\hfill
\fi
\ifnum\c@page=1
  {\sl Published in Annals of Physics, {\bf 270}, 1-51 (1998).}\hfill
\fi
} \let\@evenfoot\@oddfoot
\makeatother
\section{INTRODUCTION}

Perhaps the ultimate hope for physicists is that we will one day discover 
what is jocularly referred to as a {\it TOE}, a ``Theory of Everything", 
an all-embracing and self-consistent physical theory that summarizes 
everything that there is to know about the workings of the physical world.
Almost all physicists would undoubtedly agree that such a theory is still
conspicuous with its absence, although the agreement is probably poorer
on the issue of what would qualify as a TOE. Although the requirements that
\begin{itemize}
\item it should be a self-consistent theory encompassing quantum field 
theory and general relativity as special cases and 
\item it should have predictive power so as to be falsifiable 
in Popper's sense
\end{itemize}
are hardly controversial, it is far from clear 
which of our experimental results we should expect it to predict with 
certainty and which it should predict only in a statistical sense.
For instance, should we expect it to predict the 
masses of the elementary particles (measured in dimensionless 
Planck units, say) from first principles, are these masses 
free parameters in the theory, or do they arise from some symmetry-breaking 
process that can produce a number of distinct outcomes so that the 
TOE for all practical purposes merely predicts a probability 
distribution
\cite{Coleman,Linde 1988,Linde 1990,Vilenkin,Albrecht}?

\subsection{A classification of TOEs}

\begin{figure}[phbt]
\centerline{\rotate[r]{\vbox{\epsfysize=18cm\epsfbox{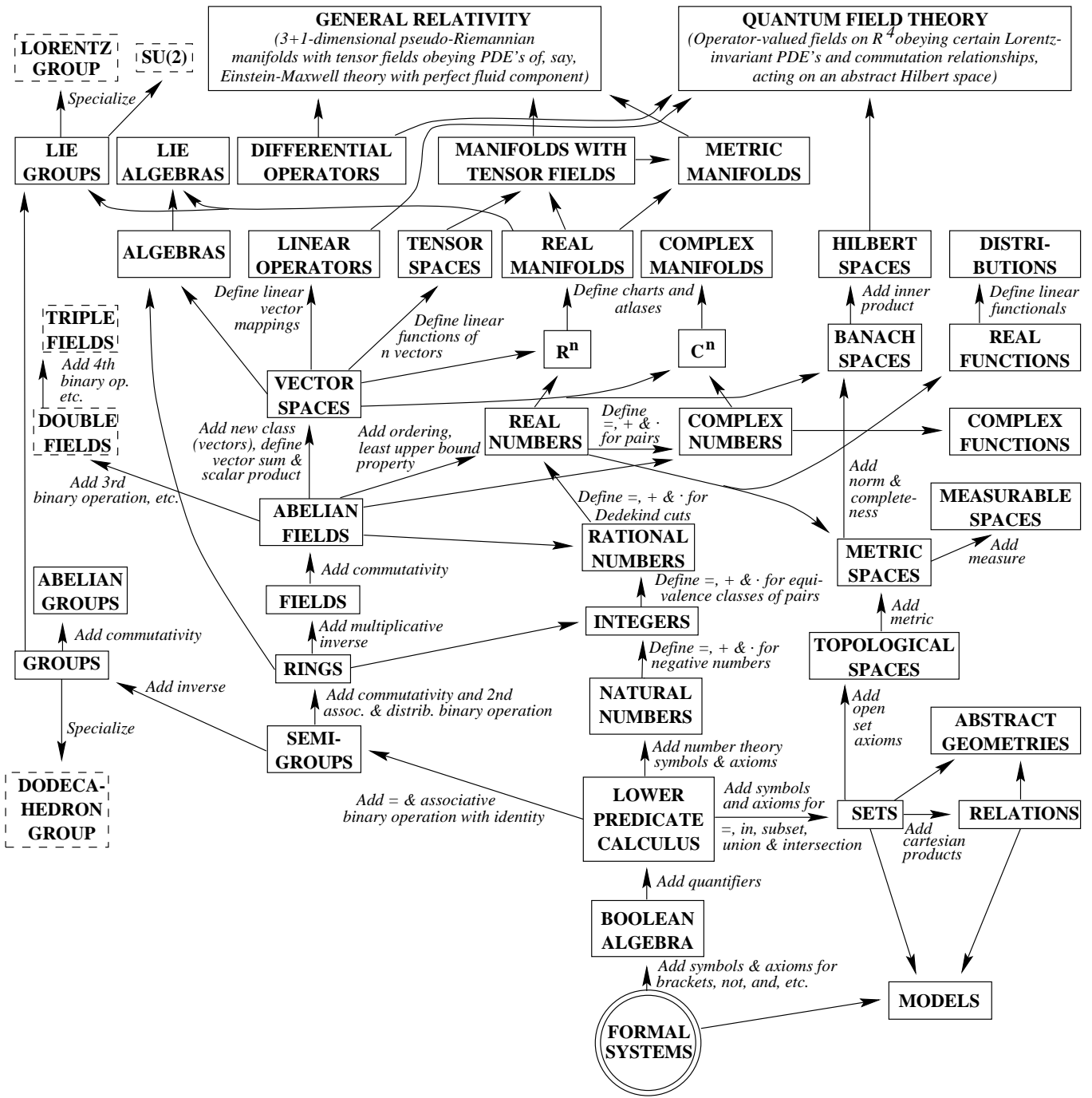}}}}
\smallskip
\caption{
\label{TreeFig}
Relationships between various basic mathematical structures.
The arrows generally indicate addition of new symbols and/or axioms.
Arrows that meet indicate the combination of structures 
--- for instance,
an algebra is a vector space that is also a ring, 
and a Lie group is a group that is also a manifold.
}
\end{figure}

Let us divide TOEs into two categories 
depending on their 
answer to the following question:
Is the physical world purely mathematical, or is mathematics merely 
a useful tool that approximately describes certain aspects of the physical
world? More formally, {\it is the physical world isomorphic to some 
mathematical structure?} 
For instance, if it were not for quantum phenomena and the problem of 
describing matter in classical theories, a tenable TOE in the first category 
would be one stating that the physical world was isomorphic to 
a 3+1-dimensional pseudo-Riemannian manifold, on which 
a number of tensor fields were defined and obeyed a certain 
system of partial differential equations. 
Thus the broad picture in a category 1 TOE is this:
\begin{itemize}
\item There are one or more mathematical structures that exist not
only in the mathematical sense, but in a physical sense as well.
\item Self-aware substructures (SASs) might inhabit some of these structures,
and we humans are examples of such SASs.
\end{itemize}
In other words, some subset of all mathematical structures 
(see \fig{TreeFig} for examples) 
is endowed with an elusive quality that we call physical existence,
or {\it PE} for brevity. Specifying this subset thus specifies 
a category 1 TOE. Since there are three disjoint possibilities
(none, some or all mathematical structures have PE), we
obtain the following classification scheme:
\begin{enumerate}
\item The physical world is completely mathematical.
\begin{enumerate}
\item Everything that exists mathematically exists physically.
\item Some things that exist mathematically exist physically,
others do not.
\item Nothing that exists mathematically exists physically.
\end{enumerate}
\item The physical world is not completely mathematical.
\end{enumerate}
The beliefs of most physicists probably 
fall into categories 2 (for instance on religious grounds) 
and 1b. Category 2 TOEs are somewhat of a resignation in the sense
of giving up physical predictive power, and will not be further 
discussed here.
The obviously ruled out 
category 1c TOE 
was only included for completeness.
TOEs in the popular category 1b are vulnerable to the
criticism (made {\eg} by Wheeler \cite{WheelerBook},
Nozick \cite{Nozick} and Weinberg \cite{Weinberg 1992}) 
that they leave an important question 
unanswered: why is that particular subset endowed with PE, 
not another? 
\onecolumn
\twocolumn
\noindent
What breaks the symmetry between mathematical 
structures that would a priori appear to have equal merit?
For instance, take the above-mentioned category 1b TOE
of classical general relativity. Then why should
one set of initial conditions have PE when other similar ones 
do not? Why should the mathematical structure where the 
electron/proton mass ratio $m_p/m_e\approx 1836$ 
have PE when the one with $m_p/m_e=1996$ does not?
And why should a 3+1-dimensional manifold have PE
when a 17+5-dimensional one does not?
In summary, although a category 1b TOE may one day turn out
to be correct, it may come to appear somewhat arbitrary
and thus perhaps disappointing to scientists hoping for a TOE 
that elegantly answers all outstanding questions and 
leaves no doubt that it really is the ultimate TOE.

In this paper, we propose that category 1a is the correct one.
This is akin to what has been termed 
{\it ``the principle of fecundity"} \cite{Nozick}, that all
logically acceptable worlds exist, 
although as will become clear
in the discussion of purely formal mathematical structures in
Section~\ref{MEsec}, the 1a TOE 
involves no difficult-to-define vestige of human language such
as ``logically acceptable" in its definition.
1a can also be viewed as a form of radical Platonism, 
asserting that the mathematical structures 
in Plato's {\it realm of ideas}, the {\it Mindscape} of Rucker
\cite{Rucker}, exist ``out there'' in a physical sense
\cite{DaviesGod}, 
akin to what Barrow refers to as ``pi in the sky''
\cite{BarrowTOE,BarrowPi}.

Since 1a is (as 1c)
a completely specified theory, involving no free parameters
whatsoever, it is in fact a candidate for a TOE.
Although it may at first appear as though 1a is just as obviously
ruled out by experience as 1c, we will argue that this is 
in fact far from clear.

\subsection{How to make predictions using this theory}

How does one make quantitative predictions using the 1a TOE?
In {\it any} theory, we can make quantitative predictions
in the form of probability distributions 
by using Bayesean statistics\footnote{
In the Bayesean view, probabilities are merely subjective
quantities, like odds, useful for making predictions.}.
For instance, to predict the classical period $T$ of a Foucault
pendulum, we would use the equation 
\beq{FocaultEq}
T= 2\pi \sqrt{L\over g}.
\eeq
Using probability distributions to model the errors
in our measurements of the
length $L$ and the local gravitational acceleration $g$,
we readily compute the probability distribution for $T$.
Usually we only care about the mean $\langle T\rangle$ and
the standard deviation $\Delta T$ (``the error bars"),
and as long as $\Delta T/T \ll 1$,
we get the mean by inserting the means in \eq{FocaultEq}
and $\Delta T/T$ from the standard expression for the 
propagation of errors.
In addition to the propagated uncertainty in the 
prior observations, the nature of the mathematical structure 
itself might add some uncertainty, as is the case in 
quantum mechanics.
In the language of the previous section, 
both of these sources of uncertainty reflect our lack
of knowledge as to which of the many 
SASs in the mathematical structure
corresponds to the one making the experiment: imperfect
knowledge of field quantities (like $g$) corresponds to uncertainty
as to where in the spacetime manifold one is, and 
quantum uncertainty stems from lack of knowledge as
to which branch of the wavefunction one is in (after the
measurement).

In the 1a TOE, there is a third source of uncertainty as 
well: we do not know exactly which mathematical structure we are 
part of, {\ie}, where we are in a hypothetical expansion of \fig{TreeFig}
containing all structures.
Clearly, we can eliminate many options as inconsistent with
our observations (indeed, many can of course be eliminated
{\it a priori} as ``dead worlds" containing no SASs at
all --- for instance, all structures in \fig{TreeFig} are
presumably too simple to contain SASs).
If we could examine all of them,
and some set of mathematical structures remained as candidates, 
then they would each make a prediction for the form 
of \eq{FocaultEq}, leaving us with a probability distribution
as to which equation to use. 
Including this uncertainty in the probability 
calculation would then give us our predicted mean and error bars.

Although this prescription may sound unfamiliar, it is quite analogous
to what we do all the time. 
We usually imagine that the fine structure constant 
$\alpha$ has some definite value in the mathematical structure
that describes our world (or at least a value where the fundamental 
uncertainty is 
substantially smaller than the measurement errors in our current 
best estimate, $1/137.0359895$). 
To reflect our measurement errors on $\alpha$, 
we therefore calculate error bars as if there were an entire 
ensemble of possible theories, spanning a small range of 
$\alpha$-values. Indeed, this Bayesean procedure
has already been applied to ensembles
of theories with radically different values of 
physical constants \cite{Vilenkin,Lambda}.
According to the 1a TOE, we must go further and include 
our uncertainty about other aspects of the mathematical structure as well, 
for instance, uncertainty as to which equations to use. 
However, this is also little different from what we do anyway,
when searching for alternative models.

\subsection{So what is new?}

Since the above prescription was found to be so similar 
to the conventional one, we must address the following question:
can the 1a TOE be distinguished from the others in practice,
or is this entire discussion merely a useless metaphysical digression?

As discussed above, the task of any theory is to 
compute probability distributions for the outcomes 
of future experiments given our previous observations.
Since the correspondence between the mathematical structure and
everyday concepts such as ``experiment" and ``outcome" 
can be quite subtle (as will be discussed at length in 
Section~\ref{PredictionSec}), it is more appropriate to rephrase this
as follows:
Given the subjective perceptions of a SAS,
a theory should allow us to compute probability distributions 
for (at least certain quantitative aspects of) its future perceptions.
Since this calculation involves summing over
{\it all} mathematical structures, 
the 1a TOE makes the following two predictions
that distinguish it from the others categories:
\begin{itemize}
\item {\bf Prediction 1:} 
The mathematical structure describing our world
is the most generic one that is consistent
with our observations.
\item {\bf Prediction 2:} 
Our observations are the most generic ones that
are consistent with our existence.
\end{itemize}
In both cases, we are referring to the totality of all observations
that we have made so far in our life. The nature of the set of 
all mathematical structures (over which we need some form 
of measure to formalize what we mean by generic) will be discussed
at length in Section~\ref{MEsec}.

These two predictions are of quite different character.
The first one offers both a useful guide when searching for the ultimate 
structure (if the 1a TOE is correct) and a way of 
predicting experimental results to potentially rule out the 1a TOE, 
as described in Section~\ref{PredictionSec}.
The second one is not practically useful, but provides 
many additional ways of potentially ruling 
the theory out.
For instance, the structure labeled ``general relativity" in
\fig{TreeFig} contains the rather arbitrary number 
3 as the dimensionality of space. Since manifolds of arbitrarily
high dimensionality constitute equally consistent mathematical 
structures, a 3-dimensional one is far from ``generic" and would
have measure zero in this family of structures.
The observation that our space appears three-dimensional would
therefore rule out the 1a TOE if the alternatives were not
inconsistent with the very existence of SASs.
Intriguingly, as discussed in Section~\ref{IslandSec}, 
all higher dimensionalities {\it do}
appear to be inconsistent with the existence of SASs, since 
among other things, they preclude stable atoms.

\subsection{Is this related to the anthropic principle?}

Yes, marginally: the weak anthropic principle must be taken into account 
when trying to rule the theory out based on prediction 2. 

Prediction 2 implies that the 1a TOE is ruled out if
there is anything about the observed Universe that is surprising, 
given that we exist. So is it ruled out?
For instance, the author has no right to be surprised 
about facts that {\it a priori} would seem unlikely, 
such as that his grandparents happened to meet or that the spermatozoid carrying
half of his genetic makeup happened to come first in a 
race agains millions of others,
as long as these facts are necessary for his existence. 
Likewise, we humans have no right to be surprised that the 
coupling constant of the strong interaction, $\alpha_s$, is not
$4\%$ larger than it is, for if it were, the sun would immediately explode
(the diproton would have a bound state, which would increase the solar 
luminosity by a factor $10^{18}$ \cite{Dyson 1971}). 
This rather tautological (but often overlooked) statement 
that we have no right to be surprised about 
things necessary for our existence has 
been termed {\it the weak anthropic principle} \cite{Carter 1974}.
In fact, investigation of the effects of varying physical parameters
has gradually revealed that \cite{Davies,BT,Balashov}
\begin{itemize}
\item
{\it virtually no physical parameters can be changed by large amounts 
without causing radical qualitative changes to the
physical world.}
\end{itemize}
In other words, the ``island" in parameter space that supports human 
life appears to be quite small.
This smallness is an embarrassment for TOEs in 
category 1b, since such TOEs provide no answer to the
pressing question of why the mathematical structures possessing
PE happen to belong to that tiny island, and has been hailed 
as support for religion-based TOEs in category 2.
Such ``design arguments" \cite{BT} stating that the world was designed by
a divine creator so as to contain SASs are closely related to 
what is termed the {\it strong anthropic principle} 
\cite{Carter 1974},
which states that the Universe must support life.
The smallness of the island has also been used to argue in favor of 
various ensemble theories in category 1b, since if structures with PE 
cover a large region of parameter space, it is not surprising  
if they happen to cover the island as well.
The same argument of course supports our 1a TOE as well, since it
in fact predicts that structures on this island (as well as all
others) have PE.

In conclusion, when comparing the merits of TOE 1a and the others,
it is important to calculate which aspects of the 
physical world are necessary for the existence of SASs and which
are not.
Any clearly demonstrated 
feature of ``fine tuning" which is unnecessary for the existence of 
SASs would immediately rule out the 1a TOE.
For this reason, we will devote Section~\ref{IslandSec}
to exploring the ``local neighborhood" of mathematical structures,
to see by how much our physical world can be changed without becoming
uninhabitable.

\subsection{How this paper is organized}

The remainder of this paper 
is organized as follows.
Section~\ref{MEsec} discusses which structures exist mathematically,
which defines the grand ensemble of which 
our world is assumed to be a member.
Section \ref{PredictionSec} discusses how to
make physical predictions using the 1a TOE.
It comments on the subtle question of how mathematical structures
are perceived by SASs in them, and proposes criteria
that mathematical structures should satisfy in order to 
be able to contain SASs.
Section~\ref{IslandSec} uses the three proposed criteria 
to map out our local island of habitability, discussing
the effects of varying physical constants, the dimensionality
of space and time, {\etc}
Finally, our conclusions are summarized in 
section \ref{ConclusionsSec}.

\section{MATHEMATICAL STRUCTURES}
\label{MEsec}

Our proposed TOE can be summarized as follows:
\begin{itemize}
\item {\it Physical existence is equivalent to mathematical existence.}\footnote{
Tipler has postulated that what he calls a 
{\it simulation} (which is similar to what we call a mathematical 
structure) has PE if and only if it contains at 
least one SAS \cite{Tipler}.
From our operational definition of PE 
(that a mathematical structure has PE if 
all SASs in it subjectively perceive themselves as existing in a physically 
real sense), it follows that the difference between this postulate and ours
is merely semantical.
}
\end{itemize}
What precisely is meant by mathematical existence, or {\it ME} for brevity?
A generally accepted interpretation of ME is that of David Hilbert:
\begin{itemize}
\item {\it Mathematical existence is merely freedom from contradiction.}
\end{itemize}
In other words, if the set of axioms that define a mathematical structure
cannot be used to prove both a statement and its negation, then 
the mathematical structure is said to have ME.

The purpose of this section is to remind the 
non-mathematician reader of the purely formal foundations of 
mathematics,  
clarifying how extensive (and limited) 
the ``ultimate ensemble" of physical worlds really is,
thereby placing Nozick's notion \cite{Nozick} of ``all logically 
acceptable worlds" on a more rigorous footing.
The discussion is centered around \fig{TreeFig}.
By giving examples,
we will illustrate the following points:
\begin{itemize}
\item The notion of a mathematical structure is well-defined.
\item Although a rich variety of structures enjoy mathematical 
existence, the variety is limited by the requirement of self-consistency
and by the identification of isomorphic ones.
\item Mathematical structures are ``emergent concepts" in a sense 
resembling that in which physical structures (say classical macroscopic 
objects) are emergent concepts in physics.
\item It appears likely that the most basic mathematical structures
that we humans have uncovered to date are the same as those that 
other SASs would find.
\item Symmetries and invariance properties are more the rule than the exception
in mathematical structures.
\end{itemize}

\subsection{Formal systems}

For a more rigorous and detailed introduction to
formal systems, the interested reader is referred to 
pedagogical books on the subject such as \cite{Hilbert,Logic}.

The mathematics that we are all taught in school is an example
of a formal system, albeit usually with rather sloppy notation.
To a logician, a {\it formal system} consists of
\begin{itemize} 
\item A collection of symbols (like for instance 
``$\icke$", ``$\imp$" and ``X")
which can be strung together into strings (like 
``$\icke\icke X\imp X$" and ``$XXXXX$")

\item 
A set of rules for determining which such strings
are {\it well-formed formulas}, abbreviated 
WFFs and pronounced ``woofs" by logicians

\item
A set of rules for determining which WFFs 
are {\it theorems}

\end{itemize}

\subsection{Boolean algebra}
\label{BooleanSec}

The formal system known as {\it Boolean algebra}
can be defined using the symbols 
``$\icke$", ``$\eller$", ``$[$", ``$]$"
and a number of 
letters ``$x$", ``$y$", ... (these letters are referred to as 
{\it variables}). 
The set of rules for determining what is a WFF are recursive:
\begin{itemize}
\item A single variable is a WFF.
\item If the strings $S$ and $T$ are WFFs, then the strings
$[\icke S]$ and $[S\eller T]$ are both WFFs.
\end{itemize}
Finally, the rules for determining what is a theorem consist of
two parts: a list of WFFs which are stated to be theorems
(the WFFs on this list are called {\it axioms}), and
rules of inference for deriving further theorems from the axioms.
The axioms are the following:
\begin{enumerate}
\item $[[x\eller x]\imp x]$
\item $[x\imp [x\eller y]]$
\item $[[x\eller y]\imp[y\eller x]]$
\item $[[x\imp y]\imp[[z\eller x]\imp [z\eller y]]]$
\end{enumerate}
The symbol ``$\imp$" appearing here is not part of the formal system.
The string ``$x\imp y$" is 
merely a convenient abbreviation for ``$[\icke x]\eller y$".
The rules of inference are
\begin{itemize}
\item The {\it rule of substitution}: 
if the string $S$ is a WFF and the string $T$ is 
a theorem containing a variable, 
then the string obtained by replacing 
this variable by $S$ is a theorem.

\item {\it Modus ponens:}
if the string $[S\imp T]$ is a theorem and 
$S$ is a theorem, then $T$ is a theorem.

\end{itemize}
Two additional convenient abbreviations are 
$``x\och y"$ (defined as ``$\icke[[\icke x]\eller[\icke y]]$"
and $``x\equiv y"$ (defined as ``$[x\imp y]\och[y\imp x]$").
Although customarily pronounced ``not", ``or",
``and", ``implies" and ``is equivalent to",  
the symbols $\icke$, $\eller$, $\och$, $\imp$ 
and $\equiv$ have no meaning
whatever assigned to them --- rather, any ``meaning" that we chose
to associate with them is an emergent concept 
stemming from the axioms and rules of inference.

Using nothing but these rules, all theorems in Boolean algebra textbooks
can be derived, from simple strings such as 
``$[x\eller[\icke x]]$" to arbitrarily long
strings.

The formal system of Boolean algebra has a number of properties that
gives it a special status among the infinitely many 
other formal systems. 
It is well-known that Boolean algebra is {\it complete}, which means 
that given an arbitrary WFF $S$, either $S$ is provable
or its negation, $[\icke S]$ is provable. 
If any of the four axioms above were removed, the system would no longer
be complete in this sense.  This means that if 
an additional axiom $S$ is added, it
must either be provable from the other axioms (and hence unnecessary)
or inconsistent with the other axioms.
Moreover, it can be shown that
``$[[x\och[\icke x]]\imp y]$" is a theorem, {\ie}, that if 
both a WFF and its negation is provable, then
{\it every WFF becomes provable}. Thus adding an independent 
(non-provable) axiom to a complete set of axioms will 
reduce the entire formal system to a banality, 
equivalent to the trivial formal system defined by 
``all WFFs are theorems"
\footnote{
Our definition of a mathematical structure having PE was that if
it contained a SAS, then this SAS would subjectively 
perceive itself as existing.
This means that Hilbert's definition of mathematical existence 
as self-consistency does not matter for our purposes, 
since inconsistent systems are too trivial to 
contain SASs anyway. Likewise, endowing ``equal rights"
to PE to formal systems below Boolean algebra, where negation 
is not even defined and Hilbert's criterion thus cannot be
applied, would appear to make no difference, since
these formal systems seem to be far too simple 
to contain SASs.
}.
Thus the formal system of Boolean algebra is not as 
arbitrary as it may at first seem. In fact, it is so basic that
almost all formal systems deemed complex enough by mathematicians
to warrant their study are obtained by starting with Boolean algebra and
augmenting it with further symbols and 
axioms.\footnote{As we saw, adding more axioms to Boolean algebra 
without adding new symbols is a losing proposition.}
For this reason, it appears at the bottom of the ``tree" of structures
in \fig{TreeFig}.

\subsection{What we mean by a mathematical structure}
\label{MSdefSec}

The above-mentioned example of Boolean algebra illustrates several 
important points about formal systems in general.
No matter how many (consistent)
axioms are added to a complete formal system, 
the set of theorems remains the same. Moreover, there
are in general a large number of different choices of independent axioms that
are equivalent in the sense that they lead to the same set of theorems,
and a systematic attempt to reduce the theorems 
to as few independent axioms as
possible would recover all of these choices independently of which one was used 
as a starting point.
Continuing our Boolean algebra example, 
upgrading $\och$, $\imp$ and $\equiv$ to 
fundamental symbols and adding additional axioms, one can again
obtain a formal system where the set of theorems remains the same.
Conversely, as discovered by Sheffer, an equivalent formal system 
can be obtained with even fewer symbols, by introducing a symbol
``$\sheffer$" and demoting ``$\icke x$" and ``$x\eller y$" to mere
abbreviations for ``$x\sheffer x$" and 
``$[x\sheffer x]\sheffer[y\sheffer y]$", respectively.
Also, the exact choice of notation is of course completely immaterial
--- a formal system with ``\%" in place of ``$\imp$" or with the bracket 
system eliminated by means of reverse Polish notation would obviously be
isomorphic and thus for mathematical purposes one and the same. 
In summary, although there are many different ways of describing 
the mathematical structure known as Boolean algebra, they are in a 
well-defined sense all equivalent. 
The same can be said about all other mathematical structures that 
we will be discussing. All formal systems can thus be
subdivided into a set of disjoint equivalence classes, 
such that the systems in each class all describe the same structure.
When we speak of a mathematical structure, 
we will mean such an equivalence class, 
{\ie}, {\it that structure which is independent of our way of describing it}.
It is to this structure that our 1a TOE attributes physical existence.

\subsection{``Mathematics space" and its limits}

If all mathematical structures have PE, then it is clearly desirable to 
have a crude overview of what mathematical structures there are.
\Fig{TreeFig} is by no means such a complete overview. Rather, it 
contains a selection of structures (solid rectangles), 
roughly based on the
Mathematics Subject Classification of the American Mathematical Society.
They have been ordered so that following an arrow 
corresponds to adding additional symbols and/or axioms.
We will now discuss some features of this ``tree" or ``web" 
that are relevant
to the 1a TOE, illustrated by examples.

\subsubsection{Lower predicate calculus and beyond}

We will not describe the WFF rules for the formal systems mentioned
below, since the reader is certainly familiar with notation such as
that for parentheses, variables and quantifiers, as well as with 
the various conventions for when parentheses and brackets can be omitted.
For instance, within the formal system of number theory described below, 
a logician would read
\beq{PrimeEq}
P(n) \equiv [(\forall a)(\forall b)[[a>1]\och[b>1] \imp a\cdot b \neq n]]
\eeq
as $P(n)$ being the statement that the natural number $n$ is prime, \ie, 
that for all natural numbers $a$ and $b$ exceeding unity,
$a\cdot b\neq n$,
and  
\beq{GoldbachEq}
[a>0] (\exists b)(\exists c)[P(b)\och P(c)\och[a+a=b+c]]
\eeq
as the (still unproven) 
Goldbach hypothesis that all even numbers can be written as the sum of two
primes. (Again, we emphasize that despite 
that we conventionally read them this way, the symbols of a formal system
have no meaning whatsoever --- the properties that we humans 
coin words for merely emerge from the axioms and rules of inference.)

Moving upward in \fig{TreeFig}, the formal system known as 
{\it Lower Predicate Calculus}
is be obtained from Boolean Algebra by adding 
quantifiers ($\forall$ and $\exists$).
One way of doing this is to add the axioms 
\begin{enumerate}
\item $[(\forall a)[A(a)]]\imp A(b)$ 
\item $[(\forall a)[A\eller B(a)]\imp [A\eller [(\forall a) B(a)]]$ 
\end{enumerate}
and the rule that if $A(a)$ is a theorem involving no quantifyers for 
$a$, then  
$(\forall a)[A(a)]$ is a theorem 
---  $(\exists a)[A(a)]$ can then be taken as a mere 
abbreviation for $\icke (\forall a)[\icke A(a)]$.

Virtually all mathematical structures
above lower predicate calculus in \fig{TreeFig}
involve the notion of {\it equality}.
The relation $a=b$ (sometimes written as 
$E(a,b)$ instead) obeys the so called axioms of equality:
\begin{enumerate}
\item $a=a$ 
\item $a=b\imp b=a$ 
\item $[a=b]\och [b=c]\imp a=c$
\item $a=b\imp [A(a)\imp A(b)]$ 
\end{enumerate}
The first three (reflexivity, symmetry and transitivity) 
make ``$=$" a so-called equivalence relation, and the last one
is known as the condition of substitutivity.

Continuing upward in \fig{TreeFig}, the formal system known as 
{\it Number Theory} (the natural numbers under addition and
multiplication) can be 
obtained from Lower Predicate Calculus by adding the
five symbols
$``=", ``0"$, $``'"$, $``+"$ and $``\cdot"$, the axioms of equality, 
and the following axioms \cite{Hilbert}: 

\bigskip
\begin{enumerate}
\item $\icke[a'=0]$
\item $[a'=b']\imp [a=b]$
\item $a+0=a$
\item $a+b' = (a+b)'$
\item $a\cdot 0=0$
\item $a\cdot b'=a\cdot b + a$
\item $[A(0)\och[(\forall a)[A(a)\imp A(a')]]]\imp A(b)$
\end{enumerate}
Convenient symbols like $1$, $2$, $``\neq"$, $``<"$, $``>"$, 
$``\leq"$ and $``\geq"$
can then be introduced as mere abbreviations  --- $1$ as an 
abbreviation for  $0'$, $2$ as an 
abbreviation for  $0''$, 
``$a\leq b$" as an abbreviation for 
$``(\exists c)[a+c=b]"$, {\etc}
This formal system is already complex enough 
to be able to ``talk about itself", 
which formally means that
G\"odels incompleteness theorem applies: there are WFFs 
such that neither they nor their negations can be proven.

Alternatively, adding other familiar axioms leads to other
branches in \fig{TreeFig}.

A slightly unusual position in the mathematical family tree
is that labeled by ``models" in the figure.
Model theory (see \eg \cite{Robinson})
studies the relationship between formal systems 
and set-theoretical models of them in terms of 
sets of objects and relations that hold between these
objects. For instance, the set of real numbers constitute 
a model for the field axioms. In the 1a TOE, all mathematical structures
have PE, so set-theoretical models of a formal system enjoy the same
PE that the formal system itself does.

\begin{figure}[phbt]
\centerline{\rotate[r]{\vbox{\epsfysize=8.8cm\epsfbox{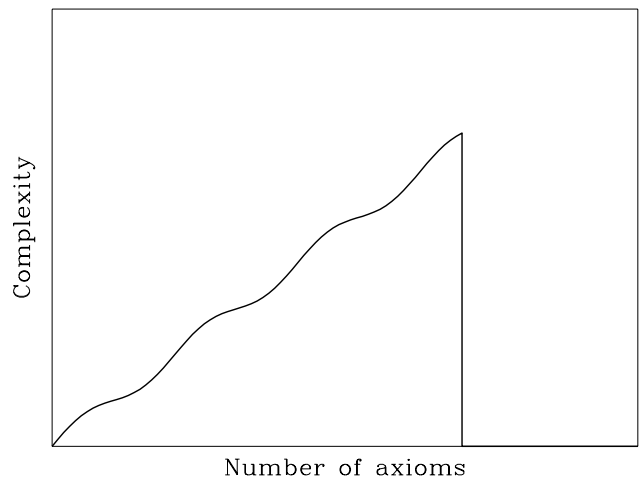}}}}
\smallskip
\caption{
\label{ComplexityFig}
With too few axioms, a mathematical structure is too simple to
contain SASs. With too many, it becomes inconsistent and thus trivial.
}
\end{figure}

\subsubsection{The limits of variety}

The variety of mathematical structures, a small part of which
is illustrated in \fig{TreeFig}, is limited in two ways.
First of all, they are much fewer than the formal systems,
since they are equivalence classes as 
described in Section~\ref{MSdefSec}. 
For instance, natural numbers (under $+$ and $\cdot$) are 
a single mathematical structure, even though
there are numerous equivalent ways of 
axiomatizing the formal system of number theory.

Second, the self-consistency requirement adds a natural
cutoff as one tries to proceed too far along arrows
in \fig{TreeFig}.
If one keeps adding axioms in the attempt to create a more 
complex structure, the bubble eventually bursts, as schematically
illustrated in \fig{ComplexityFig}. 
For instance, consider the 
progression from semi-groups to groups to finite groups
to the specific case of the dodecahedron group (a dashed 
rectangle in \fig{TreeFig}), where each successive
addition of axioms has made the structure less general
and more specific.
Since the entire multiplication table of the dodecahedron group is
specified, any attempt to make the dodecahedron
still more ``specific" will make the formal system inconsistent,
reducing it to the banal one where all WWF's are theorems.

\Fig{TreeFig} also illustrates a more subtle occurrence of such
a ``complexity cutoff", where even adding new symbols 
does not prevent a branch of the tree from ending.
Since a field is in a sense a double group 
(a group under the $1^{st}$ binary operation, and 
after removing its identity, also a group under the $2^{nd}$),
it might seem natural to explore analogous triple groups, 
quadruple groups, {\etc}
In \Fig{TreeFig}, these structures are labeled ``double fields" 
and ``triple fields", respectively. 
A double field $D$ has 3 binary operations, 
say $\Delta$, $+$ and $\cdot$, having identities that we will denote
$\infty$, $0$ and $1$, respectively, such that
$D$ under $\Delta$ and $+$, 
and $D$ without $\infty$ under $+$ and $\cdot$
are both fields. A triple field is defined analogously.
For simplicity, we will limit ourselves to ones with a finite number
of elements.
Whereas there is a rich and infinite variety of finite fields (Galois Fields), 
it can be shown \cite{Doughty} 
that (apart from a rather trivial one with 3 elements),
there is only one double field per Mersenne prime
(primes of the form $2^n-1$), and it is not known whether there are 
infinitely many Mersenne primes. As to finite triple fields,
it can be shown \cite{Doughty} that there are none at all, 
{\ie}, that particular mathematical structure in 
\fig{TreeFig} is inconsistent and hence trivial.

An analogous case of a ``terminating branch" 
is found by trying to extend the 
ladder of Abelian fields beyond the sequence rational, 
real and complex numbers.
By appropriately defining $=$, $+$ and $\cdot$ for
quadruples (rather than pairs) of real numbers,
one obtains the field of quaternions, also known as $SU(2)$.
However, the Abelian property has been lost.
Repeating the same idea for larger sets of real numbers
gives matrices, which do not even form a field
(since the sum of two invertible matrices can be
non-invertible).

\subsection{The sense in which mathematical structures are
an emergent concept}

Suppose a SAS were given the rules of some formal system and asked
to compile a catalog of theorems (for the present argument, it is 
immaterial whether the SAS is a carbon-based life-form like ourselves,
a sophisticated computer program or something completely different). 
One can then argue that
\cite{Shapiro} it would eventually invent additional
notation and branches of mathematics beyond 
this particular formal system, simply as a means of 
performing its task more efficiently. As a specific
example, suppose that the formal system is Boolean 
algebra as we defined it in Section \ref{BooleanSec},
and that the SAS tries to make a list of all strings shorter 
than some prescribed length which are theorems
(the reader is encouraged to try this).
After blindly applying the rules of inference for a while, it
would begin to recognize certain patterns and realize that
it could save considerable amounts of effort by introducing 
convenient notation reflecting these patterns. For instance, 
it would with virtual certainty introduce 
some notation corresponding 
to ``$\icke[[\icke x]\eller[\icke y]]$" (for which we 
used the abbreviation ``$\imp$") at some point, as well as 
notation corresponding what we called ``$\och$" and ``$\equiv$".
If as a starting point we had given the SAS the 
above-mentioned Sheffer version of Boolean 
algebra instead, it would surely have invented notation
corresponding to ``$\icke$" and $\eller$ as well.
How much notation would it invent?
A borderline case would be something like writing
$``x\reverseimp y"$ as an ``abbreviation" for
$``y\imp x"$, since although the symbol 
$``\reverseimp"$ is arguably helpful,
its usefulness is so marginal that it is unclear whether
it is worth introducing it at all --- indeed, most logic textbooks
do not. 

After some time, the SAS would probably discover that it's task could be
entirely automated by inventing the concept of truth tables,
where $``[x\eller[\icke x]]"$ and its negation play the roles of 
what we call ``true" and ``false".
Furthermore, when it was investigating WFFs containing long strings
of negations, {\eg},
``$\icke\icke\icke\icke\icke\icke\icke[x\eller[\icke x]]$", it might find
it handy to introduce the notion of counting and of even and odd numbers.

To further emphasize the same point, if we gave the SAS as a starting
point the more complex formal system of number theory discussed above,
it might eventually rediscover a large part of mathematics as 
we know it to aid it in proving theorems about
the natural numbers.
As is well known, certain theorems about integers
are most easily proven by employing methods that use more 
advanced concepts such as real numbers, analytic functions, {\etc} 
Perhaps the most striking such example to date is 
the recent proof of Fermat's last theorem, which employs
state-of-the-art methods of algebraic geometry
(involving, {\eg}, semistable elliptic curves)
despite the fact that the theorem itself can be phrased
using merely integers.

\subsection{What structures have we missed?}

Conversely, the systematic study of virtually any more complicated 
mathematical structure that did not explicitly involve say integers 
would lead a SAS to reinvent the integers,
since they are virtually ubiquitous.
The same could be said about most other basic mathematical structures, 
{\eg}, groups, algebras and vector spaces.
Just as it was said that 
{\it ``all roads lead to Rome"}, we are thus arguing that
{\it ``all roads lead to the integers"} and other basic 
structures --- which we therefore 
refer to as ``emergent concepts".

If this point of view is accepted, then an immediate
conclusion is that all types of SASs would arrive at 
similar descriptions of the ``mathematics tree".
Thus although one would obviously expect a certain bias towards 
being more interested in mathematics that appears relevant to
physics, {\ie}, to the particular structure in which 
the SAS resides, it would appear unlikely that any form of SASs
would fail to discover basic structures such as say Boolean Algebra, 
integers and complex numbers. 
In other words, it would appear unlikely that we humans have 
overlooked any equally basic mathematical structures.

\subsection{Why symmetries and ensembles are natural}
\label{SymmetrySec}

Above we saw that a general feature of formal systems is that 
all elements in a set are ``born equal" -- further axioms 
are needed to discriminate between them.
This means that symmetry and invariance properties 
tend to be more the rule than the exception in mathematical structures.
For instance, a three-dimensional vector space
automatically has what a physicist
would call rotational symmetry, since no preferred direction
appears in its definition. 
Similarly, any theory of physics involving the notion of a manifold
will automatically exhibit invariance under general coordinate 
transformations, simply because the very definition of a manifold
implies that no coordinate systems are privileged over any other.

Above we also saw that the smaller one wishes the ensemble
to be, the more axioms are needed. 
Defining a highly specific mathematical structure with built-in 
dimensionless numbers such as 1/137.0359895 is far from  
trivial. Indeed, it is easy to prove
that merely a
denumerable subset of all real numbers (a subset of measure
zero) can be specified by a finite number of 
axioms
\footnote{
The proof is as follows.
Each such number can be specified by a LATeX file of 
finite length that gives the appropriate axioms in 
some appropriate notation. Since each finite LATeX 
file can be viewed as a single integer 
(interpreting all its bits as as binary decimals), 
there are no more finite LATeX files than there are integers. 
Therefore there are only countably many such files and 
only countably many 
real numbers that can be specified ``numerologically'', 
\ie, by a finite number fo axioms. In short, the set of 
real numbers that we can specify at all ($\sqrt{2}$, $\pi$, 
the root of $x^7-e^x$, \etc) has Lebesgue measure zero.
}, so writing down a formal system describing 
a 1b TOE with built-in 
``free parameters" would be difficult unless these dimensionless
numbers could all be specified ``numerologically",
as Eddington once hoped.


\section{HOW TO MAKE PHYSICAL PREDICTIONS FROM THE THEORY}
\label{PredictionSec}

How does one use the Category 1a TOE to make physical predictions?
Might it really be possible that a TOE whose specification 
contains virtually no information can nonetheless 
make predictions such as that we will perceive ourselves as
living in a space with three dimensions, {\etc}?
Heretic as it may sound, we will argue that yes, 
this might really be possible, and outline a program for 
how this could be done. Roughly speaking, this involves 
examining which mathematical structures might contain SASs,
and calculating what these would subjectively appear like to the SASs that 
inhabit them. By requiring that this subjective appearance be consistent
with all our observations, we obtain a list of structures
that are candidates for being the one we inhabit.
The final result is a probability distribution
for what we should expect to perceive when we make an experiment,
using Bayesian statistics to incorporate our lack of knowledge
as to precisely which mathematical structure we reside in.

\subsection{The inside view and the outside view}

A key issue is to understand the relationship between two 
different ways of perceiving a mathematical structure. 
On one hand, there is what we will call the 
{\it ``view from outside"}, 
or the {\it ``bird perspective"}, 
which is the way in which a mathematician 
views it. 
On the other hand, there is what we will call the 
{\it ``view from inside"}, 
or the {\it ``frog perspective"}, which is the way it appears to 
a SAS in it. Let us illustrate this distinction with 
a few examples:

\subsubsection{Classical celestial mechanics}

Here the distinction is 
so slight that it is easy to overlook it altogether. 
The outside view is that of a set of vector-valued functions
of one variable, $\r_i(t)\in\R^3$, $i=1,2,...$,
obeying a set of coupled nonlinear second order ordinary differential 
equations. The inside view is that of a number of objects
(say planets and stars) at locations $\r$ in a three-dimensional 
space, whose positions are changing with time.

\subsubsection{Electrodynamics in special relativity}

Here one choice of outside view is that of a set of real-valued
functions $F_{\mu\nu}$ and $J_\mu$ in $\R^4$
which obey a certain system of 
Lorentz-invariant linear partial differential equations; 
the Maxwell equations and the equation of motion 
$(J_{\mu,\nu} - F_{\mu\nu})J^\nu = 0$.
The correspondence between this and the inside view 
is much more subtle than in the above example,
involving a number of notions that 
may seem counter-intuitive. 
Arguably, the greatest difficulty in formulating this theory
was not in finding the mathematical structure but in 
interpreting it. 
Indeed, the correct equations had to a large extent already been 
written down by Lorentz and others, but it took Albert Einstein
to explain how to relate these mathematical objects to what we 
SASs actually perceive, for instance by pointing out that 
the inside view in fact depended not only on the
position of the SAS but also on the velocity of the SAS.
A second bold idea was the notion that although the bird perspective
is that of a four-dimensional world that just is, where nothing ever
happens, it will appear from the frog perspective as a 
three-dimensional world that keeps changing.

\subsubsection{General relativity}

Here history repeated itself: although the mathematics
of the outside view had largely been 
developed by others ({\eg}, Minkowski and Riemann),
it took the genius of Einstein to 
relate this to the subjective experience from the frog
perspective of a SAS. As an illustration of 
the difficulty of relating the inside and outside views, 
consider the following scenario.
On the eve of his death, Newton was approached by 
a genie who granted him one last wish. 
After some contemplation, he made up his mind:\\
{\it ``Please tell me what the state-of-the-art equations of gravity 
 will be in 300 years."}\\
The genie scribbled down the Einstein field equations and the geodesic
equation on a sheet of paper, and being a kind genie, it also 
gave the explicit expressions for the Christoffel symbols
and the Einstein tensor in terms of the metric and 
explained to Newton how to translate the index and comma notation
into his own mathematical notation. 
Would it be obvious to Newton 
how to interpret this as a generalization of his own theory?

\subsubsection{Nonrelativistic quantum mechanics}

Here the difficulty of relating the two viewpoints reached a new
record high, manifested in the fact that 
physicists still argue about how to interpret the theory 
today, 70 years after its inception.
Here one choice of outside view is that of a Hilbert
space where a wave function evolves deterministically,
whereas the inside view is that of a world where 
things happen seemingly at random, with probability
distributions that can be computed 
to great accuracy from the wave function.
It took over 30 years from the birth of quantum mechanics
until Everett \cite{Everett} showed how the inside view
could be related with this outside view.
Discovering decoherence, which was crucial for reconciling
the presence of macrosuperpositions in the bird perspective
with their absence in the frog perspective,
came more than a decade later
\cite{Zeh 1970}.
Indeed, some physicists still find this correspondence so subtle that
they prefer the Copenhagen interpretation with a Category 2 
TOE where there is no outside view at all.

\subsubsection{Quantum gravity}
Based on the above progression of examples, one would naturally 
expect the correct theory of quantum gravity to pose even more 
difficult interpretational problems, since it must incorporate all 
of the above as special cases. 
A recent review \cite{Isham 1991} poses the following
pertinent question: {\it is the central problem of quantum gravity one of
physics, mathematics or philosophy?}
Suppose that on the eve of the next large quantum gravity meeting,
our friend the genie broke into the lecture hall and scribbled the
equations of the ultimate theory on the blackboard. 
Would any of the participants realize what was being erased the 
next morning?

\subsection{Computing probabilities}

Let us now introduce some notation corresponding to these
two viewpoints.

\subsubsection{Locations and perceptions}

Let $X$ denote what a certain 
SAS subjectively perceives at a given instant.
To be able to predict $X$, we need to specify three things:
\begin{itemize}
\item Which mathematical structure this SAS is part of
\item Which of the many SASs in this structure is this one
\item Which instant (according to the time perception of the SAS) 
we are considering.
\end{itemize}
We will label the mathematical structures by $i$, the SASs in 
structure $i$ by $j$ and the subjective time of 
SAS $(i,j)$ by $t$ --- this is a purely formal labeling, and 
our use of sums below in no way implies that these quantities
are discrete or even denumerably infinite.
We will refer to the set of all three quantities,
$L\equiv (i,j,t)$, as a {\it location}. 
If the world is purely mathematical so that
a TOE in category 1 is correct, then 
specifying the location of a perception is 
in principle sufficient to calculate what the 
perception will be. Although such a calculation is 
obviously not easy, as we will return to below, let us 
for the moment ignore this purely technical difficulty 
and investigate how predictions can be made. 

\subsubsection{Making predictions}

\begin{figure}[phbt]
\centerline{\rotate[r]{\vbox{\epsfysize=7.6cm\epsfbox{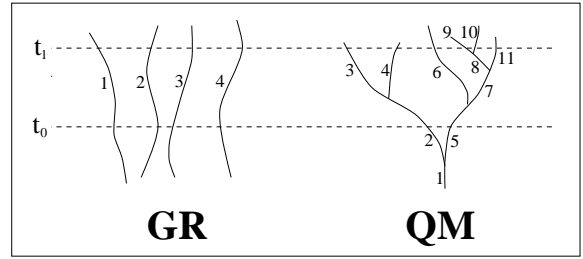}}}}
\smallskip
\caption{
\label{BranchesFig}
Whereas there is generally a 1-1 correspondence between
SASs at different times in classical general relativity 
(GR), ``observer branching" makes the situation more complicated in
quantum mechanics (QM).
}
\end{figure}

Suppose that a SAS at location $L_0$ has 
perceived $Y$ up until that instant, and that this SAS is interested
in predicting the perception $X$ that it will have
a subjective time interval $\Delta t$ into the future,
at location $L_1$.
The SAS clearly has no way of knowing {\it a priori}
what the locations $L_0$ and $L_1$ are (all it knows 
is $Y$, what it has 
perceived), so it must use statistics to reflect 
this uncertainty.
A well-known law of probability theory 
tells us that for any mutually exclusive and collectively exhaustive set of
possibilities $B_i$, the probability 
of an event $A$ is given by $P(A) = \sum_i P(A|B_i)P(B_i)$. 
Using this twice, we find that the probability
of $X$ given $Y$ is 
\beq{ProbEq1}
P(X|Y) = \sum_{L_0}\sum_{L_1}
P(X|L_1)P(L_1|L_0) P(L_0|Y).
\eeq
Since a SAS is by definition only in a single 
mathematical structure $i$ and since 
$t_1 = t_0+\Delta t$, the second factor will clearly 
be of the form
\beqa{ProbEq2}
P(L_1|L_0) 
&=& P(i_1,j_1,t_1|i_0,j_0,t_0) \nonumber\\
&=& \delta_{i_0 i_1}\delta(t_1-t_0-\Delta t) 
P(j_1|i_0,j_0).
\eeqa
If, in addition, there is a 1-1 correspondence between
the SASs at $t_0$ and $t_1$, then 
we would have simply $P(j_1|i_0,j_0) = \delta_{j_1j_2}$.
Although this is the case in, for instance, classical
general relativity, this is not the case in 
universally valid quantum mechanics, as schematically
illustrated in \fig{BranchesFig}.
As to the third factor in \eq{ProbEq1},
applying Bayes' theorem with a uniform prior for the locations
gives $P(L_0|Y)\propto P(Y|L_0)$, so 
\eq{ProbEq1} reduces to 
\beqa{ProbEq3}
&&P(X|Y) \propto \ \nonumber\\
&&\sum_{i j_0 j_1 t}
P(X|i,j_1,t+\Delta t) P(j_1|i,j_0) P(Y|i,j_0,t),
\eeqa
where $P(X|Y)$ should be normalized so as to be a probability 
distribution for $X$.
For the simple case when $P(j_1|i_0,j_0) =\delta_{j_1j_2}$,
we see that the resulting equation
\beq{ProbEq4}
P(X|Y) \propto \sum_{ijt}
P(X|i,j,t+\Delta t)P(Y|i,j,t)
\eeq
has quite a simple interpretation.
Since knowledge of a location $L=(i,j,t)$ 
uniquely determines the corresponding perception,
the two probabilities in this expression 
are either zero or unity.
We have $P(X|L)=1$ if a SAS at location $L$ perceives
$X$ and $P(X|L)=0$ if it does not.
$P(X|Y)$ is therefore simply a sum giving weight 
$1$ to all cases that are consistent with 
perceiving $Y$ and then perceiving $X$ a time interval
$\Delta t$ later, normalized so as to be a probability
distribution over $X$.
The interpretation of the more general \eq{ProbEq3}
is analogous: the possibility of ``observer branching" simply
forces us to take into account that we do not with certainty
know the location $L_1$ where $X$ is perceived 
even if we know $L_0$ exactly. 

\subsubsection{How 1a and 1b TOEs differ}

\Eq{ProbEq3} clarifies how the predictions from 
TOE 1a differ from those in 1b: 
whereas the sum should be extended over all mathematical structures
$i$ in the former case, it should be restricted to a certain subset
(that which is postulated to have PE) in the latter case.
In addition, 1b TOEs often include  
prescriptions for how to determine 
what we denote $P(X|L)$, {\ie}, the correspondence
between the inside and outside viewpoints, 
the correspondence between the numbers that we measure 
experimentally and the objects in the mathematical structure.
Such prescriptions are convenient simply because the calculation of 
$P(X|L)$ is so difficult.
Nonetheless, it should be borne in mind that this
is strictly speaking redundant information, since
$P(X|L)$ can in principle be computed {\it a priori}.

\subsection{Inside {vs.} outside: what has history taught us?}

In the past, the logical development has generally been to start with
our frog perspective and search for a bird perspective (a mathematical 
structure) consistent with it. To explore the implications of our
proposed TOE using \eq{ProbEq3}, 
we face the reverse problem: given the latter, what can we
say about the former? 
Given a mathematical structure containing a 
SAS, what will this SAS perceive, {\ie}, what is $P(X|L)$? 
This question is clearly relevant to all TOEs in category
1, including 1b. Indeed, few would dispute that a 1b TOE would gain in 
elegance if any of its postulates in the above-mentioned
``prescription" category could be trimmed away with
Occam's razor by being shown to follow from the other postulates.

Perhaps the best guide in addressing this question is what we
have (arguably) learned from previous successful theories. 
We discuss such lessons below.

\subsubsection{Do not despair}

Needless to say, the question of what a SAS would perceive in a given 
mathematical structure is a very difficult one, which we are 
far from being able to answer at the present time.
Indeed, so far we have not even found a single mathematical structure that 
we feel confident might contain SASs, since a self-consistent
model of quantum gravity remains conspicuous with its absence.
Nonetheless, the successes of relativity theory and quantum mechanics
have shown that we can make considerable progress even {\it without}
having completely solved the problem, which would of course
involve issues as difficult as understanding the human brain.
In these theories, strong conclusions were drawn about what 
could and could not be perceived that were independent of any detailed
assumption about the nature of the SAS. We list a few examples below.

\subsubsection{We perceive symmetry and invariance}

As SASs, we can only perceive those aspects of the mathematical structure that
are independent of our notation for describing it
(which is tautological, since this is all that has 
mathematical existence). For instance, as described in 
section~\ref{SymmetrySec}, if the mathematical
structure involves a manifold, a SAS can only 
perceive properties that have general covariance.

\subsubsection{We perceive only that which is useful}

We seem to perceive only those 
aspects of the mathematical structure (and of ourselves) 
that are useful to perceive, \ie, 
which are relatively stable and predictable.
Within the framework of Darwinian evolution, it
would appear as though we humans have been endowed with
self-awareness in the first place merely because certain aspects
of our world are somewhat predictable, and since this self-awareness
(our perceiving and thinking) increases our
reproductive chances\footnote{
In the 1a TOE, where all mathematical structures exist, some
SASs presumably exist anyway, without having had any evolutionary
past. Nonetheless, since processes that increase the capacity 
of SASs to multiply will have a dramatic effect on 
the various probabilities in \eq{ProbEq3}, one might expect the vast majority
of all SASs to have 
an evolutionary past. In other words, one might expect the generic 
SAS to perceive precisely those aspects the mathematical structure
which are useful to perceive. 
}.
Self-awareness would then be merely 
a side-effect of advanced information processing.
For instance, it is interesting to note that our bodily defense against
microscopic enemies (our highly complex immune system) 
does not appear to be self-aware even though our defense against
macroscopic enemies 
(our brain controlling various muscles) does.
This is presumably because the aspects of our world that
are relevant in the former case are so different
(smaller length scales, longer time scales, {\etc})
that sophisticated logical thinking and the 
accompanying self-awareness are not particularly useful here.

 Below we illustrate this usefulness criterion with three examples.

\subsubsection{Example 1: we perceive ourselves as local}

Both relativity and quantum mechanics illustrate
that we perceive ourselves as 
being ``local" even if we are not. 
Although in the bird perspective of general relativity, 
we are one-dimensional world lines
in a static four-dimensional manifold, we nonetheless 
perceive ourselves as points in a three-dimensional world where things happen.
Although a state where a person is in a superposition of two 
macroscopically different locations is perfectly legitimate 
in the bird perspective of quantum mechanics, both of these SASs will 
perceive themselves as being in a well-defined location 
in their own frog perspectives. In other words, it is only in the 
frog perspective that we SASs have a well-defined 
``local" identity at all.
Likewise, we perceive objects other than ourselves as local.

\subsubsection{Example 2: we perceive ourselves as unique}

We perceive ourselves as unique and isolated systems
even if we are not.
Although in the bird perspective of universally valid quantum mechanics, 
we can end up in several macroscopically different 
configurations at once, intricately entangled with other
systems, we perceive ourselves as remaining unique and 
isolated systems. 
What appears as ``observer branching" in the bird perspective,
appears as merely a slight randomness in the frog perspective.
In quantum mechanics, the correspondence between
these two viewpoints can be elegantly modeled with 
the density matrix formalism, where the 
approximation that we remain isolated systems corresponds to 
partial tracing over all external degrees of freedom.

\subsubsection{Example 3: we perceive that which is stable}

We human beings replace the bulk of both our 
``hardware" (our cells, say) and our ``software" (our memories,
\etc) many times in our life span. Nonetheless, we perceive ourselves as
stable and permanent \cite{Nozick}. 
Likewise, we perceive objects other than ourselves as permanent.
Or rather, what we perceive as objects are those aspects of the world
which display a certain permanence. For instance, as Eddington
remarked \cite{Eddington}, when observing the ocean 
we perceive the moving waves as objects because they
display a certain permanence, even though the water itself is only 
bobbing up and down. Similarly, we only perceive those aspects of
the world which are fairly stable to quantum 
decoherence \cite{Zeh 1970}.

\subsubsection{There can be ensembles within the ensemble}

Quantum statistical mechanics illustrates that there
can be ensembles within an ensemble. A pure state can 
correspond to a superposition of a person being in two 
different cities at once, which we, because of decoherence, 
count as two distinct SASs, but a density matrix describing a mixed
state reflects additional uncertainty as to 
which is the correct wave function.
In general, a mathematical structure $i$ may contain
two SASs that perceive themselves as belonging to 
completely disjoint and unrelated worlds.

\subsubsection{Shun classical prejudice} 

As the above discussion illustrates, 
the correspondence between the inside and outside viewpoints 
has become more subtle in each new theory (special relativity, general
relativity, quantum mechanics).
This means that we should expect it to be extremely subtle
in a quantum gravity theory and try to break all our shackles of preconception
as to what sort of mathematical structure we are looking for,
or we might not even recognize the correct equations if we see them.
For instance, criticizing the Everett interpretation of quantum mechanics
\cite{Everett} on the grounds that it is ``too crazy" would 
reflect an impermissible bias towards the familiar classical 
concepts in terms of which we humans describe our frog perspective. 
In fact, the rival Copenhagen interpretation of quantum mechanics
does not correspond to a mathematical structure at all, and 
therefore falls into category 2 \cite{Deutsch,everett}.
Rather, it attributes {\it a priori} reality to
the non-mathematical frog perspective (``the classical world") 
and denies that there is a bird perspective at all.

\subsection{Which structures contain SASs?}

Until now, we have on purpose been quite vague as to what we mean 
by a SAS, to reduce the risk of tacitly assuming that
it must resemble us humans. However, some operational definition
is of course necessary to be able to address the question in
the title of this section.
Above we asked what a SAS of a given mathematical
structure would perceive. A SAS definition allowing answers such
as ``nothing at all" or ``complete chaos" would clearly be too broad.
Rather, we picture a self-aware substructure as something capable of
some form of logical thought. We take the property of being 
``self-aware" as implicitly defined: a substructure is 
self-aware if it thinks that it is self-aware.
To be able to perceive itself as thinking (having a series
of thoughts in some logical succession), it appears as though
a SAS must subjectively perceive some form of time, either
continuous (as we do) or discrete (as our digital computers).
That it have a subjective perception of some form of space, 
on the other hand, appears far less crucial, and we will not 
require this. Nor will we insist on many of the common 
traits that are often listed in various definitions of
life (having a metabolism, ability to reproduce, {\etc}),
since they would tacitly imply a bias towards SASs similar to ours
living in a space with atoms, having a finite lifetime, {\etc}
As necessary conditions for containing a SAS, we 
will require that a mathematical structure exhibit a certain 
minimum of merely three qualities:
\begin{itemize}
\item Complexity
\item Predictability
\item Stability 
\end{itemize}
What we operationally mean by these criteria is perhaps best clarified by
the way in which we apply them to specific cases in Section~\ref{IslandSec}.
Below we make merely a few clarifying comments.
That self-awareness is likely to require the SAS 
(and hence the mathematical structure
of which it is a part) to possess a certain minimum {\it complexity} goes 
without saying. In this vein, Barrow has suggested that 
only structures complex enough for G\"odel's incompleteness 
theorem to apply can contain what we call SASs \cite{BarrowPi},
but this is of course unlikely to be a sufficient condition. 
The other two criteria are only meaningful since we required
SASs to subjectively experience some form of time.

By {\it predictability} we mean that the SAS should be able to 
use its thinking capacity to make certain inferences about its
future perceptions.
Here we include for us humans rather obvious 
predictions, such as that an empty desk is likely to remain
empty for the next second rather than, say, turn into an elephant.

By {\it stability} we mean that a SAS should exist long enough 
(according to its own subjective time) to be able to 
make the above-mentioned predictions, so this is strictly speaking 
just a weaker version of the predictability requirement.

\subsection{Why introduce the SAS concept?}

We conclude this section with a point on terminology.
The astute reader may have noticed that
the various equations in this section would have looked
identical if we have defined our observers as say ``human beings"
or ``carbon-based life-forms" instead of using the more
general term the ``SAS".
Since our prior observations $Y$ include the fact that we are
carbon-based, {\etc}, and since it is only 
Prediction 1 (mentioned in the introduction) 
that is practically useful, would it not be preferable 
to eliminate the SAS concept from the discussion altogether? 
This is obviously a matter of taste.
We have chosen to keep the discussion as general 
as possible for the following reasons:
\begin{itemize}
\item The requirement that mathematical structures should contain 
SASs provides a clean way of severely restricting the number of 
structures to sum over in \eq{ProbEq3} before getting
into nitty-gritty details involving carbon, {\etc}, which can
instead be included as part of our observations $Y$.
\item We wish to minimize the risk of anthropocentric 
and ``classical" bias when trying to establish the correspondence
between locations and perceptions. The term SAS emphasizes 
that this is to be approached as a mathematics problem, 
since the observer is nothing but part of the mathematical structure.
\item We wish to keep the discussion general enough to be applicable 
even to other possible life-forms.
\item The possibility of making {\it a priori} ``Descartes" 
predictions, based on nothing but the fact that we exist and think,
offers a way of potentially ruling out the 1a TOE that
would otherwise be overlooked.
\end{itemize}

\section{Our ``local island"}
\label{IslandSec}

In section \ref{MEsec}, we took a top-down look at
mathematical structures, starting with the most general ones
and specializing to more specific ones. In this section, we will do the
opposite: beginning with what we know about our own 
mathematical structure, we will discuss what happens when 
changing it in various ways. In other words, we will
discuss the borders of our own ``habitable island"
in the space of all mathematical structures,
without being concerned about whether there are
other habitable islands elsewhere.
We will start by discussing the effects of 
relatively minor changes such as 
dimensionless continuous physical parameters and gradually
proceed to more radical changes such as the dimensionality
of space and time and the partial differential equations.
Most of the material that we summarize in this section has been
well-known for a long time, so rather than giving a large number of
separate references, we simply refer the interested reader to 
the excellent review given in \cite{BT}. Other extensive reviews
can be found in \cite{Davies,Carr & Rees,Press & Lightman},
and the Resource Letter \cite{Balashov} contains virtually 
all references prior to 1991.

We emphasize that the purpose of this section is not to 
attempt to rigorously demonstrate 
that certain mathematical structures are devoid of SASs,
merely to provide an overview of anthropic constraints and
some crude plausibility arguments.

%
%

\subsection{Different continuous parameters}

Which dimensionless physical parameters are likely
to be important for determining the 
ability of our Universe to contain SASs?
The following six are clearly important
for low-energy physics:
\begin{itemize}
\item $\alpha_s$, the strong coupling constant 
\item $\alpha_w$, the weak coupling constant 
\item $\alpha$, the electromagnetic coupling constant 
\item $\alpha_g$, the gravitational coupling constant 
\item $m_e/m_p$, the electron/proton mass ratio
\item $m_n/m_p$, the neutron/proton mass ratio
\end{itemize}
We will use Planck units,
so $\hbar=c=G=1$ are not parameters, and 
masses are dimensionless, making the observed parameter
vector 
\beqa{ParameterVectorEq}
\nonumber
&(\alpha_s,\alpha_w,\alpha,\alpha_g,m_e/m_p,m_n/m_p) 
\approx\\ 
&(0.12,0.03,1/137,5.9\tento{-39},1/1836,1.0014).
\eeqa
Note that $m_p$ is not an additional parameter, since
$m_p=\alpha_g^{1/2}$.
In addition, the cosmological constant \cite{Lambda}
and the various 
neutrino masses will be important if they are
non-zero, since they can contribute to the overall 
density of the universe and therefore affect both its expansion rate
and its ability to form cosmological large-scale structure such as 
galaxies. How sensitive our existence is to the values of 
various high-energy physics parameters (\eg, the 
top quark mass) is less clear, but we can certainly
not at this stage rule out the possibility that their 
values are constrained by various early-universe phenomena.
For instance, the symmetry breaking that lead to a slight excess
of matter over anti-matter was certainly crucial for
the current existence of stable objects.

\begin{figure}[phbt]
\centerline{{\vbox{\epsfxsize=8.8cm\epsfbox{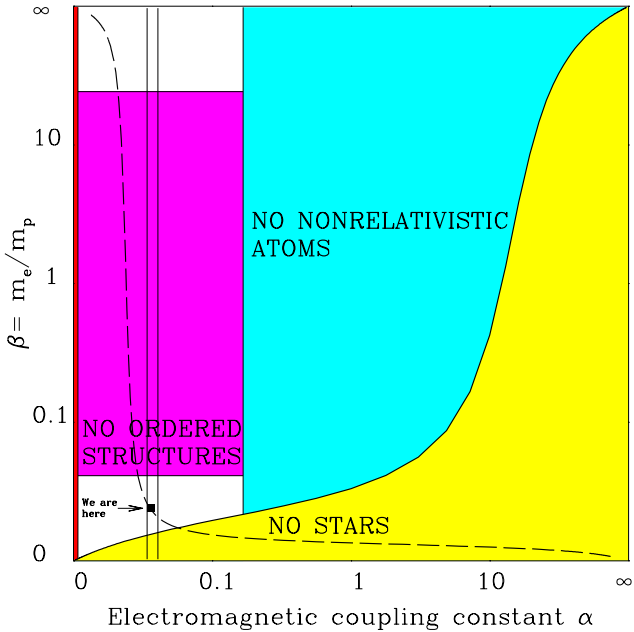}}}}
\smallskip
\caption{
\label{alphabetaFig}
Constraints on $\alpha$ and $\beta$
}
Various edges of our ``local island" are illustrated in the parameter
space of the fine structure constant $\alpha$ and the 
electron/proton mass ratio $\beta$. 
The observed values $(\alpha,\beta)\approx(1/137,1/1836)$ 
are indicated with a filled square. Grand unified theories
rule out everything except the narrow strip between the two vertical lines, 
and Carter's stellar argument predicts a point on the
dashed line. In the narrow shaded region to the very left, 
electromagnetism is weaker than gravity and 
therefore irrelevant.

\end{figure}

As was pointed out by Max Born, 
given that $m_n\approx m_p$, 
the gross properties of all atomic and molecular systems are controlled by
only two parameters: $\alpha$ and $\beta\equiv m_e/m_p$.
Some rather robust constraints on these parameters are 
illustrated by the four shaded regions in
\fig{alphabetaFig}.
Here we have compactified the parameter space by plotting 
$\arctan[\lg(\beta)]$ against 
$\arctan[\lg(\alpha)]$, 
where the logarithms are in base 10, 
thus mapping the entire range $[0,\infty]$ 
into the range $[-\pi/2,\pi/2]$. 
First of all, the fact that $\alpha, \beta\ll 1$ 
is crucial for chemistry as we know it.
In a stable ordered structure (\eg, a chromosome),
the typical fluctuation in the location of 
a nucleus relative to the inter-atomic spacing
is $\beta^{1/4}$, so for such a structure to remain
stable over long time scales, one must have
$\beta^{1/4}\ll 1$. The figure shows the rather 
modest constraint $\beta^{1/4}<1/3.$
(This stability constraint also allows replacing $\beta$ by $1/\beta$, which interchanges the roles of electrons and nucleons.)
In contrast, if one tried to build up ordered 
materials using the strong nuclear force
one would not have this important stability, 
since neutrons and protons have similar masses 
(this is why neither is localized with precision in nuclei,
which all appear fairly spherically symmetric from the
outside) \cite{BT}.

The typical electron velocity in a hydrogen atom
is $\alpha$, so $\alpha\ll 1$ makes small 
atoms nonrelativistic and atoms and molecules 
stable against pair creation.

The third constraint plotted 
is obtained if we require the existence of stars.
Insisting that the lower limit on the stellar mass 
(which arises from requiring the central temperature 
to be high enough to ignite 
nuclear fusion) not exceed the upper limit 
(which comes from requiring that the star be stable) \cite{BT}, 
the dependence on $\alpha_g$ cancels out and one 
obtains the constraint $\beta\simgt \alpha^2/200$.

A rock-bottom lower limit is clearly $\alpha\simgt\alpha_g$,
since otherwise electrical repulsion would always be weaker than
gravitational attraction, and the effects of electromagnetism would for all 
practical purposes be negligible. 
(More generally, if we let any parameter approach zero or infinity, 
the mathematical structure clearly loses complexity.) 

If one is willing to make more questionable assumptions,
far tighter constraints can be placed. For instance, 
requiring a grand-unified theory to unify all forces 
at an energy no higher than the Planck energy and requiring 
protons to be stable on stellar timescales gives 
$1/180\simlt\alpha\simlt 1/85$, the two vertical lines 
in the figure. If it is true that a star must
undergo a convective phase before arriving at the main sequence
in order to develop a planetary system, then one obtains 
\cite{Carter 1974}
the severe requirement $\alpha^{12}/\beta^4\sim\alpha_g$, 
which is plotted (dashed curve) using the
observed value of $\alpha_g$. 

In addition to the above-mentioned constraints, a
detailed study of biochemistry reveals that
many seemingly vital processes hinge on a large number 
of ``coincidences"\cite{BT},
ranging from the fact that water attains its maximum density 
above its freezing point to various chemical properties 
that enable high-fidelity DNA reproduction.
Since all of chemistry is essentially determined
by only two free parameters, $\alpha$ and $\beta$,
it might thus appear as though there 
is a solution to an overdetermined problem with much more 
equations (inequalities) than unknowns.
This could be taken as support for a religion-based
category 2 TOE, with the argument that it would be 
unlikely in all other TOEs.
An alternative interpretation is that these constraints are
rather weak and not necessary conditions for the existence
of SASs. In this picture, there would be a large number of
tiny islands of habitability in the white region in \fig{alphabetaFig},
and the SASs in other parts of this ``archipelago"
would simply evolve by combining 
chemical elements in a manner different from ours.

\begin{figure}[phbt]
\centerline{{\vbox{\epsfxsize=8.8cm\epsfbox{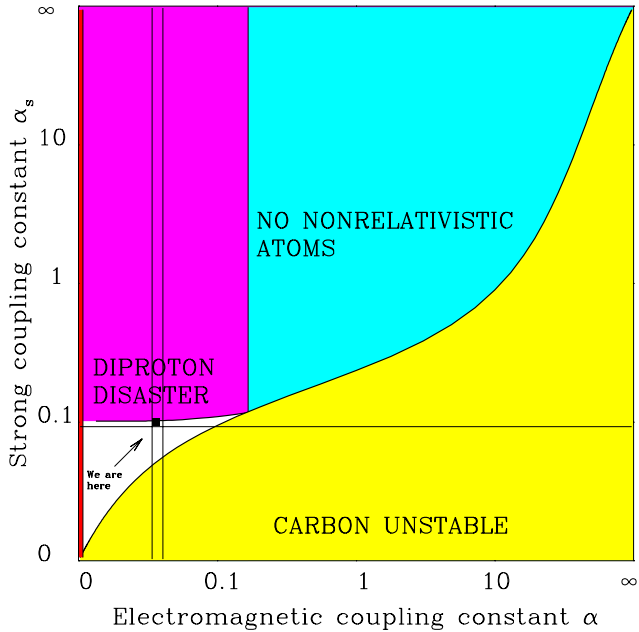}}}}
\smallskip
\caption{
\label{alphaalphasFig}
Constraints on $\alpha$ and $\alpha_s$
}
Various edges of our ``local island" are illustrated in the parameter
space of the electromagnetic and strong coupling constants,
$\alpha$ and $\alpha_s$.
The observed values $(\alpha,\alpha_s)\approx(1/137,0.1)$ 
are indicated with a filled square. Grand unified theories
rule out everything except the narrow strip between the two vertical lines, 
and deuterium becomes unstable below the horizontal line.
In the narrow shaded region to the very left, 
electromagnetism is weaker than gravity and 
therefore irrelevant.
\end{figure}

A similar constraint plot involving the strength of the
strong interaction is shown in \Fig{alphaalphasFig}.
Whether an atomic nucleus is stable or not depends on whether
the attractive force between its nucleons is able to
overcome degeneracy pressure and Coulomb repulsion.
The fact that the heaviest stable nuclei 
contain $Z\sim 10^2$ protons is therefore determined
by $\alpha_s$ and $\alpha$. 
If $\alpha_s\simlt 0.3\alpha^{1/2}$ (a shaded region), 
not even carbon $(Z=5)$ would be stable \cite{BT}, 
and it is doubtful whether the universe
would still be complex enough to support SASs.
Alternatively, increasing $\alpha_s$ by a mere 
$3.7\%$ is sufficient to endow the diproton 
with a bound state \cite{Davies 1972}.
This would have catastrophic consequences for stellar 
stability, as it would accelerate hydrogen 
burning by a factor of $10^{18}$. In fact, 
this would lead to a universe devoid of Hydrogen 
(and thus free of water and organic chemistry), 
since all $H$ would be converted to diprotons already
during big bang nucleosynthesis.
Reducing $\alpha_s$ by $11\%$ (horizontal line)
would unbind deuterium, without which the main
nuclear reaction chain in the sun could not proceed.
Although it is unclear how necessary this is for SASs, 
``it is doubtful if stable, long-lived stars could exist at
all" \cite{Davies}.

Just as in \fig{alphabetaFig}, we might expect a careful
analysis of the white
region to reveal a large number of tiny habitable islands.
This time, the source of the many delicate
constraints is the hierarchical process in which 
heavy elements are produced in stars, where it is sufficient
to break a single link in the reaction chain
to ruin everything.
For instance, Hoyle realized that
production of elements heavier than $Z=4$ (Beryllium) 
required a resonance in the $C^{12}$ nucleus 
at a very particular energy, and so predicted
the existence of this resonance before it had 
been experimentally measured \cite{Hoyle 1953}.
In analogy with the ``archipelago" argument above,
a natural conjecture is that if $\alpha_s$ and
$\alpha$ are varied by large enough amounts, alternative
reaction chains can produce enough heavy elements
to yield a complex universe with SASs.

Since it would be well beyond the scope
of this paper to enter into a detailed discussion of
the constraints on all continuous parameters, we merely list
some of the most robust constraints below and refer
to \cite{BT,Davies} for details.
\begin{itemize}
\item $\alpha_s$: See \fig{alphaalphasFig}. 
\item $\alpha_w$: 
If substantially smaller, 
all hydrogen gets converted to helium shortly after the Big Bang.
If much larger or much smaller, the neutrinos from a supernova
explosion fail to blow away the outer parts of the star,
and it is doubtful whether any heavy elements would 
ever be able to leave the stars where they were produced.
\item $\alpha$: See \fig{alphabetaFig} and \fig{alphaalphasFig}.
\item $\alpha_g$: The masses of planets, organisms on their surface, and
atoms have the ratios 
$(\alpha/\alpha_g)^{3/2}:(\alpha/\alpha_g)^{3/4}:1$,
so unless $\alpha_g\ll\alpha$, the hierarchies of scale
that characterize our universe would be absent.
These distinctions between micro- and macro- may be necessary
to provide stability against statistical $1/\sqrt{N}$-fluctuations,
as pointed out by Schr\"odinger after asking ``Why are atoms
so small?" \cite{Erwin}. Also note the Carter constraint in 
\fig{alphabetaFig}, which depends on $\alpha_g$.
\item $m_e/m_p$: See \fig{alphabetaFig}.
\item $m_n/m_p$: If $m_n/m_p<1+\beta$, then neutrons cannot decay 
into protons and electrons, so nucleosynthesis converts
virtually all hydrogen into helium. 
If $m_n/m_p<1-\beta$, then protons would be able
to decay into neutrons and positrons, so there would be
no stable atoms at all.
\end{itemize}


\subsection{Different discrete parameters}

\subsubsection{Different number of spatial dimensions}

\begin{figure}[phbt]
\centerline{{\vbox{\epsfysize=8.8cm\epsfbox{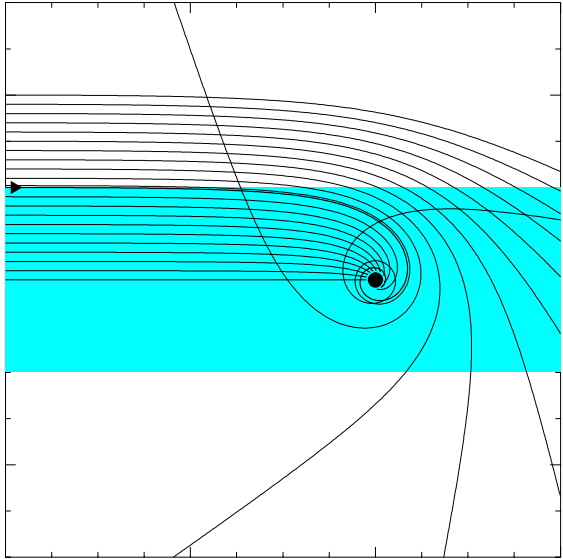}}}}
\smallskip
\caption{
\label{PlanetsFig}
The two body problem in four-dimensional space: 
the light particles that approach the heavy one at the center
either escape to infinity or get sucked into a cataclysmic collision.
There are no stable orbits.
}
\end{figure}

In a world with the same laws of physics as ours but
where the dimensionality
of space $\n$ was different from three, it it quite plausible that
no SASs would be possible.\footnote{This discussion of
the dimensionality of spacetime is an expanded 
version of \cite{dimensions}.}
What is so special about $\n=3$?
Perhaps the most striking property was pointed out by Ehrenfest 
in 1917 \cite{Ehrenfest 1917,Ehrenfest 1920}: 
for $\n>3$, neither classical atoms nor
planetary orbits can be stable. Indeed, as described below,
quantum atoms cannot be stable either.
These properties are related to the fact that the 
fundamental Green function of the Poisson equation 
$\nabla^2\phi=\rho$, which gives the electrostatic/gravitational
potential of a point particle, is $r^{2-\n}$ for $\n>2$. 
Thus the inverse square law of electrostatics and gravity 
becomes an inverse cube law if $\n=4$, {\etc}
When $\n>3$, the two-body problem no longer has any stable orbits
as solutions \cite{Buechel 1963,Freeman 1969}. 
This is illustrated in \fig{PlanetsFig},
where a swarm of light test particles are incident from
the left on a massive point particle (the black dot),
all with the same momentum vector but 
with a range of impact parameters.
There are two cases: those that start outside the shaded region
escape to infinity, whereas those with smaller impact parameters
spiral into a singular collision in a finite time.
We can think of this as there being a finite cross section for
annihilation. 
This is of course in stark contrast to 
the familiar case $\n=3$, which gives either stable elliptic orbits or 
non-bound parabolic and hyperbolic orbits, and has no 
``annihilation solutions" except for the measure zero case
where the impact parameter is exactly zero.
A similar disaster occurs in quantum mechanics, where a study 
of the Schr\"odinger equation shows that the Hydrogen 
atom has no bound states for $\n>3$ \cite{Tangherlini 1963}. 
Again, there is a finite
annihilation cross section, which is reflected by the fact 
that the Hydrogen atom has no ground state, but time-dependent 
states of arbitrarily negative energy.
The situation in general relativity is analogous
\cite{Tangherlini 1963}.
Modulo the important caveats mentioned below,
this means that such a world cannot contain any objects that
are stable over time, and thus almost certainly cannot contain
SASs.

There is also a less dramatic way in 
which $\n=3$ appears advantageous (although perhaps not necessary) 
for SASs. Many authors (\eg, 
\cite{Ehrenfest 1917,Ehrenfest 1920,Poincare 1917,Courant & Hilbert})
have drawn attention to the fact that
the wave equation admits sharp and distortion-free signal propagation
only when $\n=3$ (a property that 
we take advantage of when we see and listen).
In addition, the Weyl equation exhibits a special 
``form stability'' \cite{NielsenRugh} for $\n=3$, $\m=1$.

\begin{figure}[phbt]
\centerline{{\vbox{\epsfxsize=8.8cm\epsfbox{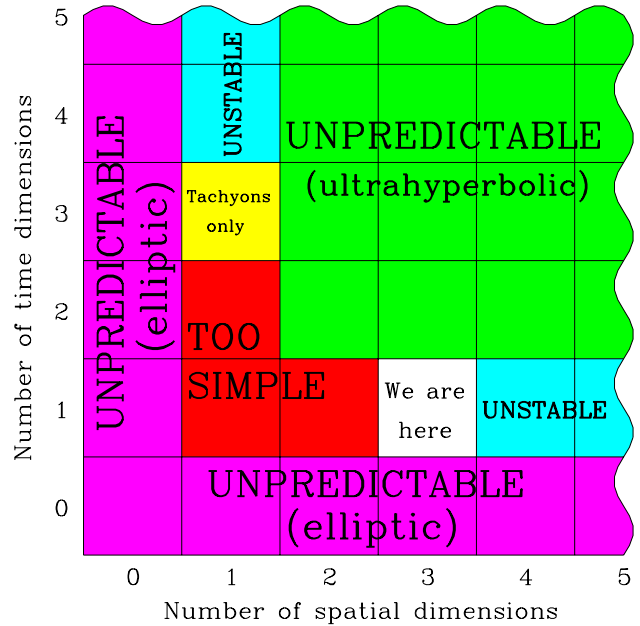}}}}
\smallskip
\caption{
\label{DimensionsFig}
Constraints on the dimensionality of spacetime.
}
When the partial differential equations are elliptic or 
ultrahyperbolic, physics has no predictive power for a SAS.
In the remaining (hyperbolic) 
cases, $n>3$ fails on the stability requirement
(atoms are unstable) and $\n<3$ fails on the 
complexity requirement (no gravitational attraction, topological
problems).
A 1+3-dimensional spacetime is equivalent to a 3+1-dimensional
one with tachyons only, and may fail on the stability requirement.
\end{figure}

What about $\n<3$? 
It has been argued \cite{Whitrow 1955} that organisms
would face insurmountable topological problems if $\n=2$: 
for instance, two nerves cannot cross.
Another problem, emphasized by Wheeler \cite{MTW}, 
is the well-known fact 
(see {\eg} \cite{Deser 1984}) that there 
is no gravitational force in general relativity with
$\n<3$.
This may appear surprising, since the Poisson equation of 
Newtonian gravity allows gravitational forces for $\n<3$.
The fact of the matter is that general relativity has 
no Newtonian limit for $\n<3$, which means that a naive
Newtonian calculation would simply be inconsistent with
observations.
In order for general relativity to possess a Newtonian limit, 
there must be a coordinate system where the 
metric tensor $g_{\mu\nu}$ approaches the Minkowski form
far from any masses, and the Newtonian gravitational potential
$\phi$ is then (apart from a factor of two) identified with the deviation
of $g_{00}$ from unity.
However, a naive application of Newtonian gravity shows that
the inverse $r^2$ law of 
$\n=3$ gets replaced by an inverse $r$ law for $\n=2$, so instead
of approaching zero as $r\to\infty$,
$\phi$ diverges logarithmically.
A general relativistic solution for a static point particle 
in 2 dimensional space (which is mathematically equivalent to
the frequently studied problem of mass concentrated on an 
infinite line, interpreted as a cosmic string) shows that
the surrounding space has no curvature (the Riemann tensor vanishes),
which means that other particles would not feel any gravitational
attraction. Instead, the surrounding spacetime has the geometry of 
a cone, so that the total angle around the point mass is less
than $360^\circ$, again illustrating the global nature of 
gravity in two dimensions and why there cannot be a Newtonian limit.
In summary, there will not be any gravitational attraction 
if space has less than 3 dimensions, which precludes planetary orbits
as well as stars and planets held together by gravity
and SASs being gravitationally bound to rotating planets.
We will not spend more time listing problems with $\n<3$, but simply 
conjecture that since 
$\n=2$ (let alone $\n=1$ and $\n=0$) offers vastly less
complexity than $\n=3$, worlds with $\n<3$ are just too
simple and barren to contain SASs.
  
For additional discussion of anthropic constraints on $\n$, 
see {\eg} \cite{BT,Barrow 1983}.

\subsubsection{Different number of time dimensions}

Why is time one-dimensional?
In this section, we will present an argument for why a  
world with the same laws of physics as ours and
with an $\n+\m$-dimensional spacetime can only
contain SASs if the number of time-dimensions $\m=1$, 
regardless of the number of space-dimensions $\n$.
Before describing this argument, which 
involves hyperbolicity properties of partial differential 
equations, let us make a few general comments about 
the dimensionality of time.

Whereas the case $\n\ne 3$ has been frequently discussed in the literature,
the case $\m\ne 1$ has received rather scant attention.
This may be partly because the above-mentioned correspondence between 
the outside and inside viewpoints is more difficult to establish in
the latter case. When trying to 
imagine life in 4-dimensional space, we can make an analogy
with the step from a 2-dimensional world to our
3-dimensional one, much as was done in Edwin Abbot's famous
novel ``Flatland". 
But what would reality appear like to a SAS in a manifold with
two time-like dimensions?

A first point to note is that even for $\m>1$, there is no 
obvious reason for 
why a SAS could not nonetheless {\it perceive} 
time as being one-dimensional, thereby 
maintaining the pattern of having ``thoughts" and
``perceptions" in the one-dimensional succession
that characterizes our own reality perception.
If a SAS is a localized object, it will travel along 
an essentially 1-dimensional (time-like) world line
through the $\n+\m$-dimensional spacetime manifold. The  
standard general relativity notion of its proper time is 
perfectly well-defined, and we would expect this to be the time that
it would measure if it had a clock and that it would subjectively
experience.

\paragraph{Differences when time is multidimensional}

Needless to say, many aspects of the world would nonetheless
appear quite different. For instance, 
a re-derivation of relativistic mechanics for this more 
general case shows that energy now becomes an 
$\m$-dimensional vector rather than a constant, whose
direction determines in which of the many time-directions
the world-line will continue, and in the non-relativistic
limit, this direction is a constant of motion. In other 
words, if two non-relativistic observers that are moving in 
different time directions happen to meet at a point in spacetime,
they will inevitably drift apart in separate
time-directions again, unable to stay together. 

Another interesting difference, which can be shown by an elegant 
geometrical argument \cite{Dorling 1969}, is that
particles become less stable when $\m>1$. 
For a particle to be able to decay when $\m=1$, 
it is not sufficient that there 
exists a set of particles with the same quantum numbers.
It is also necessary, as is well-known, that the sum of their rest 
masses should be less than the rest mass of the original 
particle, regardless of how great its kinetic energy may be.
When $\m>1$, this constraint vanishes \cite{Dorling 1969}.
For instance, 
\begin{itemize}
\item 
a proton can decay into a neutron,  
a positron and a neutrino, 
\item
an electron can decay into a neutron, an antiproton and a neutrino,
and 
\item
a photon of sufficiently high energy can decay 
into any particle and its antiparticle.
\end{itemize}
In addition to these two differences, one can
concoct seemingly strange occurrences involving 
``backward causation" when $\m>1$.
Nonetheless, although such unfamiliar behavior may appear disturbing, 
it would seem unwarranted to assume that it would prevent any
form of SAS from existing. After all, we must avoid the fallacy
of assuming that the design of our human bodies
is the only one that allows self-awareness. 
Electrons, protons and photons would still be stable if
their kinetic energies were low enough, so perhaps observers could
still exist in rather 
cold regions of a world with $m>1$\footnote{
It is, however, far from trivial to formulate a quantum field theory with a
stable vacuum state when $m>1$. 
A detailed discussion of such instability
problems with $\m>1$ is given by Linde \cite{Linde 1990},
also in an anthropic context, 
and these issues are closely related to the
ultrahyperbolic property described below.
}
There is, however, an additional problem for SASs 
when $\m>1$, which has not been
previously emphasized even though the mathematical results
on which it is based are well-known.
It stems from the requirement of {\it predictability} which was discussed in
Section~\ref{PredictionSec}.\footnote{
The etymology of {\it predict} makes it a slightly unfortunate
word to use in this context, since it might be interpreted as presupposing
a one-dimensional time. We will use it to mean merely making 
inferences about other parts of the space-time manifold based on local data.
}
If a SAS is to be able to make any use of its self-awareness
and information-processing abilities (let alone function),
the laws of physics must be such that
it can make at least some predictions. Specifically, 
within the framework of a field theory, it should by 
measuring various nearby field values be able to
compute field values at some more distant space-time points 
(ones lying along its future world-line being particularly useful)
with non-infinite error bars.
Although this predictability requirement may sound modest, 
it is in fact only met by a small class of partial differential 
equations (PDEs), essentially those which are hyperbolic.

\paragraph{The PDE classification scheme}

All the mathematical material summarized below is well-known,
and can be found in more detail in \cite{Courant & Hilbert}.
Given an arbitrary second order linear partial differential equation
in $\R^d$,
\beq{PDEeq}
\left[\sum_{i=1}^d \sum_{j=1}^d 
A_{ij} {\partial\over\partial x_i} {\partial\over\partial x_j}
+ \sum_{i=1}^d b_i {\partial\over\partial x_i}
+ c\right] u=0,
\eeq
where the matrix $A$ (which we without loss of generality 
can take to be symmetric), the vector $b$ and the scalar $c$
are given differentiable functions of the $d$ coordinates,
it is customary to classify it  
depending on the signs of the eigenvalues of $A$. The PDE is 
said to be
\begin{itemize}
\item
{\it elliptic} in some region of $\R^d$ if they are all positive 
or all negative there,
\item {\it hyperbolic} if one is positive and the rest are 
negative (or vice versa), and
\item {\it ultrahyperbolic} in the remaining case, {\ie},  
where at least two eigenvalues are positive and
at least two are negative.
\end{itemize}
What does this have to do with the dimensionality of spacetime?
For the various covariant field equations of nature that describe
our world (the wave equation $u_{;\mu\mu}=0$, 
the Klein-Gordon equation $u_{;\mu\mu} + m^2 u=0$, \etc\footnote{
Our discussion will apply to matter fields with spin as well,
{\eg} fermions and photons,
since spin does not alter the causal structure of the solutions.
For instance, all four components of an electron-positron field 
obeying the Dirac equation satisfy the Klein-Gordon equation as well,
and all four components of the electromagnetic vector potential 
in Lorentz gauge satisfy the wave equation. 
}), the matrix $A$ will clearly have the same eigenvalues 
as the metric tensor.
For instance, they will be hyperbolic in a metric of 
signature $(+---)$, corresponding to $(\n,m)=(3,1)$,
elliptic in a metric of signature $(+++++)$,
corresponding to $(\n,m)=(5,0)$, and ultrahyperbolic in
a metric of signature $(++--)$.

\paragraph{Well-posed and ill-posed problems}

One of the merits of this standard 
classification of PDEs is that it determines their causal structure,
\ie, how the boundary conditions must be specified to make the
problem {\it well-posed}.
Roughly speaking, the problem is said to be well-posed if 
the boundary conditions determine a unique solution $u$ 
and if the dependence of this solution on the boundary
data (which will always be linear) is {\it bounded}.
The last requirement means that the solution $u$ at a given point 
will only change by a finite amount if the boundary data
is changed by a finite amount.
Therefore, even if an ill-posed problem can be formally 
solved, this solution would in practice be useless to a SAS,
since it would need to measure the initial data with infinite
accuracy to be able to place finite error bars on the solution
(any measurement error would cause the error bars on the solution
to be infinite).

\paragraph{The elliptic case}

Elliptic equations allow well-posed {\it boundary value problems}.
For instance, the $d$-dimensional 
Laplace equation $\nabla^2 u=0$ 
with $u$ specified on 
some closed $(d-1)$-dimensional 
hypersurface determines the solution everywhere inside
this surface. On the other hand, giving ``initial" data 
for an elliptic PDE on a non-closed surface, say a plane, 
is an ill-posed problem.\footnote{
Specifying only $u$ on a non-closed surface
gives an underdetermined problem, and specifying additional data,
{\eg}, the normal derivative of $u$, generally makes the problem
over-determined
and ill-posed in the same way as the ultrahyperbolic case described below.
}
This means that
a SAS in a world with no time dimensions (\m=0) would
not be able do make any inferences at all about the situation
in other parts of space based on what it observes locally.
Such worlds thus fail on the above-mentioned predictability
requirement, as illustrated in \fig{DimensionsFig}.
 
\paragraph{The hyperbolic case}

\begin{figure}[phbt]
\centerline{{\vbox{\epsfysize=8.8cm\epsfbox{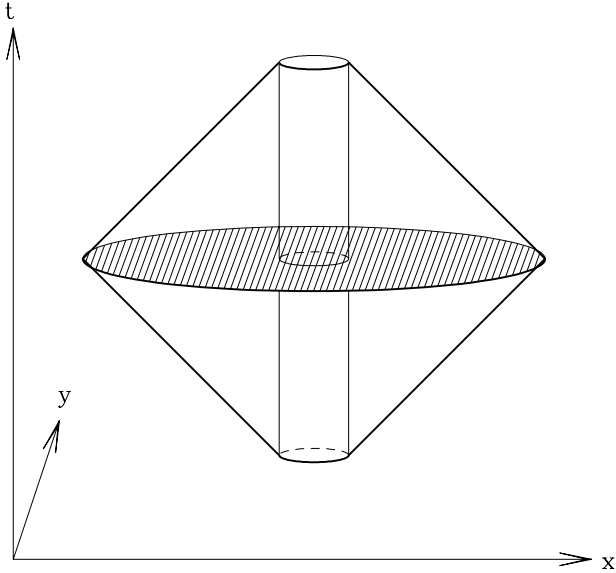}}}}
\smallskip
\caption{
\label{CausalityFig}
The causality structure for hyperbolic and ultra-hyperbolic equations.}
\end{figure}

Hyperbolic equations, on the other hand, 
allow well-posed {\it initial-value problems}.
For the Klein-Gordon equation in $\n+1$ dimensions,
specifying initial data ($u$ and $\dot u$) on
a region of a spacelike hypersurface 
determines $u$ at all points for which this region slices 
all through the backward light-cone, as long as 
$m^2\ge 0$ --- we will return to the Tachyonic case 
$m^2<0$ below.
For example, initial data on the shaded disc in \fig{CausalityFig}
determines the solution in the volumes bounded by the
two cones, including the (missing) tips.
A localized SAS can therefore make predictions about its
future. If the matter under consideration is 
of such low temperature that it is nonrelativistic,
then the fields will essentially contain only Fourier 
modes with wave numbers $|{\bf k}|\ll m$,
which means that for all practical purposes, the solution
at a point is determined by the initial data in a 
``causality cone" with an opening angle much narrower than
$45^\circ$. For instance, when we find ourselves
in a bowling alley where no relevant macroscopic
velocities exceed 10 m/s, we can use information from 
a spatial hypersurface  
of 10 meter radius (a spherical volume) to make predictions
an entire second into the future.

\paragraph{The hyperbolic case with a bad hypersurface}

In contrast, if the initial data for a hyperbolic PDE is 
specified on a hypersurface that is not spacelike, the problem 
becomes ill-posed. \Fig{CausalityFig}, which is based on
\cite{Courant & Hilbert}, provides an intuitive 
understanding of what goes wrong.
A corollary of a remarkable theorem by Asgeirsson
\cite{Asgeirsson} is that if we specify $u$ in 
the cylinder in \fig{CausalityFig}, then this determines
$u$ throughout the region made up of the truncated double
cones. Letting the radius of this cylinder approach zero, 
we obtain the disturbing conclusion 
that providing data in a for all practical purposes
one-dimensional region determines the solution in a 
three-dimensional region. 
Such an apparent ``free lunch", where the solution seems to
contain more information than the input data,
is a classical symptom of ill-posedness. The price that must be
paid is specifying the input data with infinite accuracy,
which is of course impossible given real-world measurement 
errors.\footnote{
A similar example occurs if all we know about a mapping $f$
in the complex plane is 
that it is an analytic function. Writing $z=x+iy$
and $f(x+iy) = u(x,y) + iv(x,y)$, where $x$, $y$, $u$ and
$v$ are real, 
the analyticity requirement corresponds to the Cauchy-Riemann 
PDEs 
\beq{CauchyRiemannEq}
{\partial u\over \partial x} = {\partial v\over \partial y},\quad\quad
{\partial u\over \partial y} = -{\partial v\over \partial x}.
\eeq
It is well-known that knowledge of $f$ in an infinitesimal
neighborhood of some point (for all practical purposes
a zero-dimensional region) uniquely determines $f$ everywhere
in the complex plane. However, this important 
mathematical property would be practically useless for a SAS
at the point trying to predict its surroundings, since  
the reconstruction is ill-posed: it must measure 
$f$ to infinitely many decimal places to be able
to make any predictions at all, since its Taylor series requires
knowledge of an infinite number of 
derivatives.
}
Moreover, no matter how narrow
we make the cylinder, the problem is always over-determined, 
since data in the outer half of the cylinder are 
determined by that in the inner half. 
Thus measuring data in a larger region does not eliminate the 
ill-posed nature of the problem, since the additional data 
carries no new information.
Also, generic boundary data allows no solution at all, since it is not 
self-consistent. 
It is easy to see that 
the same applies when specifying ``initial" 
data on part of a non-spacelike hypersurface,
\eg, that given by $y=0$.
These properties are analogous in $\n+1$-dimensions, and illustrate
why a SAS in an $\n+1$-dimensional spacetime can
only make predictions in time-like directions.

\paragraph{The ultrahyperbolic case}

Asgeirsson's theorem applies to the ultrahyperbolic case as well,
showing that initial data on a hypersurface containing both spacelike and 
timelike directions leads to an ill-posed problem.
However, since a hypersurface by definition has 
a dimensionality which is one less than that of the 
spacetime manifold (data on a submanifold of lower dimensionality
can never give a well-posed problem), 
{\it there are no spacelike or timelike hypersurfaces} in the
ultrahyperbolic case, {\ie}, when the number of space- and
time-dimensions both exceed one.
In other words, worlds in the region labeled ultrahyperbolic
in \fig{DimensionsFig} cannot contain SASs if we 
insist on the above-mentioned predictability requirement.
\footnote{The only remaining possibility is the rather contrived case
where data are specified on a null hypersurface.
To measure such data, a SAS would need to ``live on the light cone", 
{\ie}, travel with the speed of light, which 
means that it would subjectively not 
perceive any time at all (its proper time would stand still).
}
Together with the above-mentioned complexity and stability requirements, 
this rules out all combinations $(\n,\m)$ in \fig{DimensionsFig}
except $(3,1)$. We see that what makes the number 1 so special
is that a hypersurface in a manifold has a dimensionality
which is precisely 1 less than that of the manifold itself
(with more than one time-dimension, a hypersurface cannot
be purely spacelike).

\paragraph{Space-time dimensionality: summary}

Here we have discussed only linear PDEs, although the full system of coupled
PDEs of nature is of course non-linear. This in no way weakens our
conclusions about only $\m=1$ giving well-posed initial 
value problems. When PDEs give ill-posed problems
even {\it locally}, in a small neighborhood of a hypersurface 
(where we can generically approximate 
the nonlinear PDEs with linear ones), it is obvious that 
no nonlinear terms can make them well-posed in a larger neighborhood.
Contrariwise, adding nonlinear terms occasionally makes
well-posed problems ill-posed.

We summarize this section as follows, graphically illustrated
in \fig{DimensionsFig}. 
In our 1a TOE, there are mathematical structures with PE 
that have exactly our laws of physics but with different space-time 
dimensionality. It appears likely that all except the 
3+1-dimensional one are devoid of SASs, for the following 
reasons:
\begin{itemize}
\item More or less that 1 time dimension:\\
insufficient predictability.
\item More than 3 space dimensions:\\
insufficient stability.
\item Less than 3 space dimensions:\\
insufficient complexity.
\end{itemize}
Once again, these arguments are of course not to be interpreted as a
rigorous proof.
For instance, within the context of specific models, 
one might consider exploring the possibility of stable structures
in the case $(n,m)=(4,1)$ based
on short distance quantum corrections to the $1/r^2$ potential or on
string-like (rather than point-like) particles\cite{Witten}.
We have simply argued that it is far from obvious that any other combination 
than $(n,m)=(3,1)$ permits SASs, since radical qualitative
changes occur when $\n$ or $\m$ is altered.

\subsubsection{Including tachyonic particles}

If spacetime were 1+3-dimensional instead of 
3+1-dimensional, space and time would effectively have interchanged 
their roles, except that $m^2$ in a Klein-Gordon equation 
would have its sign reversed.
In other words, a 1+3-dimensional world would be just like ours
except that all particles would be tachyons, as illustrated in
\fig{DimensionsFig}.

Many of the original objections towards tachyons have 
been shown to be unfounded \cite{Feinberg 1967}, but
it also appears premature to conclude that 
a world with tachyons could provide SASs with the necessary
stability and predictability. 
Initial value-problems are still well-posed when data is given on a spacelike
hypersurface, but new instabilities appear.
A photon of arbitrary energy could decay into 
a tachyon-antitachyon pair \cite{Dorling 1969}, and the other 
forbidden decays that we discussed above 
(in the context of multidimensional time) 
would also become allowed.
In addition, fluctuations in the Tachyon field
of wavelength above $1/m$ would
be unstable and grow exponentially rather than oscillate.
This growth occurs on a timescale $1/m$, so if our Universe had
contained a Tachyon field with $m\simgt 1/10^{17}$ seconds, it would have
dominated the cosmic density and caused the Universe to recollapse
in a big crunch long ago.
This is why the $(\n,\m)=(1,3)$ box
is tentatively part of the excluded region in 
\fig{DimensionsFig}.

\subsection{Other obvious things to change}
 
Many other obvious small departures from our island of 
habitability remain to be better explored in this framework, 
for instance:
\begin{itemize}
\item 
Changing the spacetime topology, either on cosmological
scales or on microscopical scales.
\item
Adding and removing low-mass particles (fields).
\item 
Adding tachyonic particles (fields), as touched upon above.
\item
Making more radical changes to the partial differential equations.
The notion of hyperbolicity has been generalized also to PDEs or
order higher than two, and also in these cases we would expect
the predictability requirement to impose a strong constraint. 
Fully linear equations (where all fields are uncoupled) presumably
lack the complexity necessary for SASs, whereas non-linearity 
is notorious for introducing instability and unpredictability (chaos). 
In other words, it is not implausible that there exists only a small
number of possible systems of PDEs that balance
between violating the complexity constraint on one hand and 
violating the predictability and stability constraints on the other hand. 

\item Discretizing or $q$-deforming spacetime.
                                                                               
\end{itemize}

%
%
%
%
%
%
%
%
%
%

\clearpage 
\section{Conclusions}
\label{ConclusionsSec}

We have proposed a ``theory of everything" (TOE) which is 
in a sense the ultimate ensemble theory.
In the Everett interpretation of nonrelativistic quantum mechanics, 
many macroscopically different states of the universe are 
all assumed to have physical existence (PE), 
but the structure of spacetime, the
physical constants and the physical laws 
are assumed to be fixed. Some quantum gravity conjectures 
\cite{Hawking topology}
endow not only different metrics but even 
different spacetime topologies with PE. It
has also been suggested that physical constants may be different 
``after" a big crunch and a new big bang 
\cite{Wheeler 1977}, thus endowing 
models with say different coupling constants and particle masses 
with PE, and this same ensemble has been postulated to have
PE based on a wormhole physics argument \cite{Coleman}.
It has been argued that models with different 
effective space-time dimensionality have PE, 
based on inflationary cosmology \cite{Linde 1988} or 
superstring theory \cite{Albrecht}. The ``random dynamics"
program of Nielsen \cite{Nielsen} has even endowed PE to worlds
governed by a limited class of different equations, corresponding to
different high-energy Lagrangeans.
Our TOE takes this ensemble enlargement to its extreme,
and postulates that all structures that
exist in the mathematical sense (as described in Section~\ref{MEsec}) 
exist in the physical sense as well.
The elegance of this theory lies in its extreme simplicity, since it contains
neither any free parameters nor any arbitrary assumptions about
which of all mathematical equations are assumed to be ``the real ones".

The picture is that some of these mathematical structures 
contain ``self-aware substructures"
(SASs), and that we humans are an example of such SASs.
To calculate the physical predictions of the theory, we therefore
need to address the following questions:
\begin{enumerate}
\item Which structures contain SASs?
\item How do these SASs perceive the structures that they are part of?
\item Given what we perceive, which mathematical structures are most likely
to be the one that we inhabit?
\item Given specific experimental observations (perceptions),
what are the probability distributions for experimental outcomes
when using Bayesean statistics to reflect our lack of knowledge
(as to which structure we inhabit, as to measurement errors, {\etc})?
\end{enumerate}
Needless to say, these are all difficult questions, and an exhaustive
answer to any one of them would of course 
be far beyond the scope of a single paper.
For example, many person-years have already been spent on investigating
whether string theory is a candidate under item 3.
In this paper, we have merely attempted to give an introductory discussion
of these four questions, and we summarize our findings 
under the four headings below.

\subsection{Which structures contain SASs?}

As robust necessary conditions for the 
existence of SASs, we proposed three criteria:
\begin{itemize}
\item Complexity
\item Predictability
\item Stability 
\end{itemize}
The last two are clearly only meaningful for SAS that ``think" and thus 
subjectively perceive some form of time.
Using the terminology of Carter \cite{Carter 1974}, we are asking how large 
the ``cognizable" part of the grand ensemble is.
In Section \ref{MEsec}, we set an upper limit on its size
by exploring the ensemble of all mathematical structures, thereby 
placing Nozick's notion \cite{Nozick} of ``all logically 
acceptable worlds" on a more rigorous footing.
We noted that if one keeps adding additional axioms to a formal 
system in an attempt
to increase its complexity, one generically reaches a point where
the balloon bursts: the formal system becomes inconsistent,
all WFFs become theorems, and the mathematical structure becomes
trivial and loses all its complexity.

In Section \ref{IslandSec}, we replaced this ``top-down" approach with a
``bottom-up" approach, making an overview of our local neighborhood
in ``mathematics space". The constraints summarized here were all 
from previously published work except for a few new
observations regarding the
dimensionality of time and space.

In the six-dimensional space spanned by the 
low-energy-physics parameters
$\alpha_s$, $\alpha_w$, $\alpha$, $\alpha_g$,
$m_e/m_p$ and $m_n/m_p$, we found that 
an ``archipelago" picture emerged when 
assuming that the existence of SASs requires a certain minimum
complexity, predictability and stability.
As has been frequently emphasized, the local ``island of habitability" 
to which our world belongs is quite small, extending only over relative 
parameter variations of order $10^{-2}$.
However, since the number of constraints for our own particular
existence is much greater than the number of free parameters,
we argued that it is likely that there is an archipelago of many such 
small islands, corresponding to different nuclear reaction chains in 
stellar burning and different chemical compositions of the SASs. 
The presence of a smaller number of much more severe constraints
indicates that this archipelago also has an end, so that large regions
on parameter space are likely to be completely devoid of SASs, and it is 
likely that the total number if islands is finite. 

In the discrete two-dimensional space corresponding to different numbers
of space and time dimensions, all but the combination 3+1 appear to be
``dead worlds", 
devoid of SASs. If there were more or less than one time-dimension, 
the partial differential equations of nature would lack the hyperbolicity
property that enables SASs to make predictions.
If space has a dimensionality exceeding three, there are no atoms or
other stable structures.
If space has a dimensionality of less than three, it is doubtful
whether the world offers sufficient complexity to support SASs
(for instance, there is no gravitational force). 

We concluded that the requirements of complexity, 
predictability and stability are extremely restrictive in our ``local
neighborhood" of mathematical structures, so it is 
not implausible that that islands of habitability are small and rare 
elsewhere in ``mathematics space" as well.
For this reason, it is not obvious that there is  
more than a finite number of mathematical structures 
containing SASs that are perceptibly different (according to their 
SASs), so that it might even be possible to 
catalogue all of them.

Many other obvious small departures from our island of 
habitability remain to 
be better explored in this framework, 
for instance 
changing the spacetime topology on sub-horizon scales, 
adding and removing low-mass
particles, adding tachyonic particles, 
making more radical 
changes to the partial differential equations
and discretizing or $q$-deforming spacetime.

\subsection{How do SASs perceive the structures that they are part of?}

In Section~\ref{PredictionSec}, we argued that the 
development of relativity theory and quantum mechanics has taught
us that we must carefully distinguish between two different views
of a mathematical structure:
\begin{itemize}
\item
The {\it bird perspective} or {\it outside view}, which is the way a mathematician views it.
\item
The {\it frog perspective} or {\it inside view}, which is the way it is perceived by
a SAS in it.
\end{itemize}
Understanding how to predict the latter from the former is one of the major
challenges in working out the quantitative predictions of our proposed TOE
--- and indeed in working out the quantitative predictions of {\it any}
physical theory which is based on mathematics.
Perhaps the best guide in addressing this question is what we
have (arguably) learned from previous successful theories:

\begin{itemize}

\item 
The correspondence between the two viewpoints 
has become more subtle in each new theory (special relativity, general
relativity, quantum mechanics), so we should expect it to be extremely subtle
in a quantum gravity theory and try to break all our shackles of preconception
as to what sort of mathematical structure we are looking for.
Otherwise we might not even recognize the correct equations if we see them.

\item 
We can only perceive those aspects of the mathematical structure that
are independent of our notation for describing it. 
For instance, if the mathematical
structure involves a manifold, a SAS can only 
perceive properties that have general covariance.

\item 
We seem to perceive only those 
aspects of the structure (and of ourselves) 
which are useful to perceive, \ie, 
which are relatively stable and predictable
(this is presumably because our design is related to
Darwinian evolution).

\begin{itemize}
\item 
Example 1:
We perceive ourselves as ``local" even if we are not. 
Although in the bird perspective of general relativity
we are one-dimensional world lines
in a static four-dimensional manifold, we perceive ourselves
as points in a three-dimensional world where things happen.

\item 
Example 2:
We perceive ourselves as stable and permanent even if we are not. 
(We replace the bulk of both our hardware 
(cells) and software (memories) many times in our lifetime).

\item 
Example 3: We perceive ourselves as unique and isolated systems
even if we are not.
Although in the bird perspective of universally valid quantum mechanics 
we can end up in several macroscopically different 
configurations at once, intricately entangled with other
systems, we perceive ourselves as remaining unique isolated
systems merely experiencing a slight randomness.
\end{itemize}

\item
There can be ensembles within the ensemble:
even within a single mathematical structure (such as quantum mechanics),
different SASs can perceive different and for all practical purposes
independent physical realities. 

\end{itemize}

\subsection{Which mathematical structure do we inhabit?}

This question can clearly only be addressed by continued physics 
research along conventional lines, although we probably need to
complement mathematical and experimental 
efforts with more work on understanding 
the correspondence between the inside and outside viewpoints.

Since some aspects of complex mathematical structures can often 
be approximated by simpler ones, we might never be able 
to determine precisely which one we are part of.
However, if this should turn out to be the case, it clearly
will not matter, since we can then obtain 
all possible physical predictions
by just assuming that our structure is the simplest of
the candidates.

\subsection{Calculating probability distributions}

All predictions of this theory take the form of probability distributions
for the outcomes of future observations, as formally given
by \eq{ProbEq3}. 
This equation basically states that 
the probability distribution for observing $X$ 
a time $\Delta t$ after observing $Y$ is 
a sum giving equal weight to all mathematical structures
that are consistent with both $X$ and $Y$.
As mentioned in the introduction, this is in fact
quite similar to how predictions are made with 1b TOEs, the difference 
being simply that the sum is extended over {\it all} mathematical structures
rather than just a single one or some small selection.
It is convenient to discard all mathematical structures that
do obviously not contain SASs once and for all, as
this greatly cuts down the number of structures to sum over. 
This is why it is useful to identify and weed out 
``dead worlds" as was done in section~\ref{IslandSec}. 
After this first cut, all additional observations about the
nature of our world of course provide additional cuts in the number
of structures to sum over.

\subsection{Arguments against this theory}

Here we discuss a few obvious objections to the proposed theory.

\subsubsection{The falsifiability argument}

Using Popper's falsifiability requirement, one might argue that
{\it ``this TOE does not qualify as a scientific theory, 
since it cannot be experimentally ruled out".}
In fact, a moment of consideration reveals that this argument is false.
The TOE we have proposed makes a large number of statistical predictions,
and therefore {\it can} eventually be ruled out at high confidence levels if
it is incorrect, using prediction 1 from the introduction as
embodied in \eq{ProbEq3}.
Prediction 2 from the introduction offers additional ways of
ruling it out that other theories lack.
Such rejections based on a single observation are 
analogous to those involving statistical
predictions of quantum mechanics:
Suppose that we prepare a silver atom with its spin in the $z$-direction and
then measure its spin in a direction making an angle 
of $\theta=3^\circ$ with the $z$-axis.
Since the theory predicts the outcome to be ``spin up" with a probability
$\cos^2\theta/2$, a single observation of ``spin down" would imply that 
quantum mechanics had been ruled out at a confidence level
exceeding 99.9\%.\footnote{
\label{3Dfootnote}
Indeed, the author once thought that this TOE was
ruled out by the observation that space is three-dimensional, since 
if all higher dimensionalities were at least as habitable for SASs as our own,
the a priori probability of our living in a space so near the bottom of the
list would be virtually zero. The theory is saved merely because 
of Ehrenfest's observation that $\n>3$ precludes
atoms, which the author was unaware of at the time (and independently
rediscovered, just as many others have).
}
Instead, the falsifiability argument can be applied against
rival TOEs in category 1b, as discussed in 
Section~\ref{NonexistenceSec} below.

\subsubsection{The pragmatism argument}

One might argue that
{\it ``this TOE is useless in practice, since it cannot
make any interesting predictions".}
This argument is also incorrect.
First of all, if only one mathematical structure should turn
out to be consistent with all our observations, then we will
with 100\% certainty know that this is the one we inhabit, and
the 1a TOE will give identical predictions to the 1b TOE
that grants only this particular structure PE.
Secondly, probabilistic calculations as to which structure we inhabit
can also provide quantitative predictions. 
It was such Bayesean reasoning that enabled Hoyle to predict 
the famous 7.7 MeV resonance in the $C^{12}$ 
nucleus \cite{Hoyle 1953}, and we would expect 
the derived probability distributions to be 
quite narrow in other cases as well when parameters appear in
exponentials.
Although many in principle interesting calculations
(such as the high-resolution version of 
\fig{alphaalphasFig} suggested above) are numerically very difficult
at the present stage, 
there are also areas where this type of calculations
do not appear to be unfeasibly difficult. 
The parameters characterizing cosmological initial conditions 
provide one such example where work has already been done
\cite{Lambda,Vilenkin}. 
Shedding more light on the question of 
whether or not particle physicists should expect 
a ``mass desert" up to near the Planck scale might also be possible
by a systematic study of the extent to which ``generic" models
are consistent with our low-energy observations \cite{Nielsen}.

When it comes to discrete parameters, our TOE in fact makes some 
strikingly specific {\it a priori} predictions given the other
laws of physics, such as that our spacetime 
should be 3+1-dimensional (see \fig{DimensionsFig}).
Thus an extremely prodigal new-born baby could in principle, before 
opening its eyes for the first time, paraphrase Descartes by saying
``{\it cogito, ergo space is three-dimensional}".

\subsubsection{The economy argument}
\label{EconomySec}

One might argue that
this TOE is vulnerable to Occam's razor,
since it postulates the existence of other worlds that 
we can never observe.
{\it Why should nature be so wasteful and indulge in such opulence 
as to contain an infinite plethora of different worlds?}

Intriguingly, this argument can be turned around to argue
{\it for} our TOE. When we feel that nature is wasteful according to this
theory, what precisely
are we disturbed about her wasting? Certainly not ``space",
since we are perfectly willing to accept a single Friedmann-Robertson-Walker
Universe with an infinite volume, most of which we can never observe
--- we accept that this unobservable space has PE since it allows a 
simpler theory.
Certainly not ``mass" or ``atoms", for the same reason --- once you have 
wasted an infinite amount of something, who cares if you waste some more?
Rather, it is probably the apparent reduction in simplicity that 
appears disturbing, the quantity of information necessary to specify 
all these unseen worlds. 
However, as is discussed in more detail in \cite{nihilo},
{\it an entire ensemble is often much simpler than one of its members},
which can be stated more formally using the notion of 
{\it algorithmic information content} \cite{Solomonoff,Chaitin},
also referred to as {\it algorithmic complexity}.
%
%
For instance, the algorithmic information in a number 
is roughly speaking defined as the length (in bits) 
of the shortest computer program
which will produce that number as output, so the information 
content in a generic integer $n$ is of order $\log_2 n$.
Nonetheless, the set of all integers $1, 2, 3, ...$
can be generated by quite a 
trivial computer program \cite{ZurekAlgo1}, so the algorithmic
complexity of the whole set is smaller than that of a generic member.
Similarly, the set of all perfect fluid solutions to the 
Einstein field equations has a smaller algorithmic complexity than
a generic particular solution, since the former is specified simply by 
giving a few equations and the latter requires the specification of 
vast amounts of initial data on some hypersurface.
Loosely speaking, the apparent information content rises when
we restrict our attention to one particular element in an ensemble,
thus losing the symmetry and simplicity that was inherent in the totality
of all elements taken together.
In this sense, our ``ultimate ensemble" of all mathematical structures 
has virtually no algorithmic complexity at all.
Since it is merely in the frog perspective, in the subjective perceptions
of SASs, that this opulence of information and complexity is really there,
one can argue that an ensemble theory is in fact
more economical than one endowing only a single ensemble element 
with PE 
\cite{nihilo}.

\subsection{Arguments against the rival theories}

Here we discuss some objections to rival TOEs.
The first two arguments are against those in Category 2,
and the second two against those in Category 1b.

\subsubsection{The success of mathematics in the physical sciences}

In a his famous essay
``The Unreasonable Effectiveness of Mathematics in the
Natural Sciences" \cite{Wigner},
Wigner argues that ``the enormous usefulness of mathematics in the 
natural sciences is something
bordering on the mysterious", and that 
``there is no rational explanation for it".
This can be used as an argument against
TOEs in category 2.
For 1a and 1b TOEs, on the other hand, 
the usefulness of mathematics for describing the physical
world is a natural consequence of the fact that the latter {\it is}
a mathematical structure.
The various approximations that constitute our current physics theories
are successful because simple mathematical
structures can provide good approximations 
of how a SAS will perceive more complex mathematical structures.
In other words, our successful theories are 
not mathematics approximating physics, 
but mathematics approximating mathematics!

In our category 1a TOE, Wigner's question about why there are laws of physics
that we are able to discover follows from the above-mentioned
predictability requirement.
In short, we can paraphrasing Descartes: {\it ``Cogito, ergo lex."}

\subsubsection{The generality of mathematics}

A second challenge for defenders of a Category 2 TOE is 
that mathematics is far more than the study of numbers that
is taught in school. The currently popular formalist definition 
of mathematics as the study of formal systems 
\cite{Curry 1951} (which
is reflected in our discussion of mathematical structures 
in Section~\ref{MEsec}) is so broad that for all practical purposes,
{\it any} TOE that is definable in purely formal terms
(independent of vague human terminology)
will fall into Category 1 rather than 2. For instance, 
a TOE involving a set of different types of entities 
(denoted by words, say) and
relations between them (denoted by additional words) 
is nothing but what mathematicians call
a set-theoretical model, and one can generally find a 
formal system that it is a model of.
Since proponents of a Category 2 TOE must argue that 
some aspects of it are {\it not} mathematical,
they must thus maintain that the world is in some sense not 
describable at all. 
Physicists would thus be wasting their time looking for 
such a TOE, and one can even argue about whether such a TOE deserves 
to be called a theory in the first place.

\subsubsection{The smallness of our island}

If the archipelago of habitability 
covers merely an extremely small fraction of our local
neighborhood of ``mathematics space", then 
a Category 1b TOE would provide no explanation for the
``miracle" that the parameter values of the existing world
happen to lie in the range allowing SASs.
For instance, if nuclear physics calculations were to show that
stellar heavy element production requires 
$1/137.5 <\alpha < 1/136.5$ and a 1b TOE would 
produce a purely numerological
calculation based on gravity quantization 
giving $\alpha=1/137.0359895$, then 
this ``coincidence" would probably leave many physicists feeling 
disturbed. Why should the existence of life arise 
from a remarkable feat of fine-tuning on the part of nature?
However, in all fairness, we must bear in mind that 
extreme smallness of the archipelago (and indeed even extreme smallness
of our own island) has not been convincingly 
demonstrated, as stressed by {\eg} 
\cite{Greenstein & Kropf}.
Even the dimensionality observation mentioned in 
footnote \ref{3Dfootnote} hardly qualifies as such a fine-tuning 
argument, since a small integer appearing from a 
calculation would appear far less arbitrary than an enormous
integer or a number like 137.0359895.
Rather, we mention this fine tuning issue simply to 
encourage actual calculations of how small the islands are.

\subsubsection{Physical nonexistence is a scientifically meaningless concept}
\label{NonexistenceSec}

The claim that some mathematical structure 
(different from ours) does {\it not} have physical existence
is an empirically completely useless statement,
since it does not lead to any testable predictions, not even
probabilistic ones.
Nonetheless, it is precisely such claims that all TOEs in 
category 1b must make to distinguish themselves from the
1a TOE.
This makes them vulnerable to Popper's criticism of not qualifying
as physical theories, since this aspect of them is not falsifiable.
As an example, let us suppose that a detailed calculation shows that
only five tiny islands in \fig{alphaalphasFig} allow heavy element production
in stars, with a total area less than $10^{-6}$ of the total in the plot.
In a 1b TOE predicting the observed values, this would {\it not} allow
us to say that the theory had been ruled out with $99.9999\%$ significance, 
nor at any significance level at all, simply because, contrary to the 
1a TOE,
there is no ensemble from which probabilities emerge.
However uncomfortable we might feel about the seeming 
``miracle" that the theory happened to predict parameter values allowing 
life, a defender of the theory could confidently ignore our unease, 
knowing that the claimed nonexistence of worlds with other parameter
values could never be falsified.

Moreover, purely philosophical arguments about whether 
certain mathematical structures 
have PE or not (which is the sole difference between 1a and 1b TOEs)
appear about as pointless as the medieval purely philosophical arguments about 
whether there is a God or not, 
when we consider that the entire notion of PE is painfully
poorly defined. Would most physicists attribute PE to galaxies outside
of our horizon volume? To unobservable branches of the wavefunction?
If a mathematical structure contains a SAS, then the claim that it has PE 
operationally means that this SAS will perceive itself as existing in 
a physically real world, just as we do\footnote{
We could eliminate the whole notion of PE from our TOE by simply rephrasing 
it as {\it ``if a mathematical structure contains a SAS, it will 
perceive itself as existing in a physically real world"}.
}.
For the many other mathematical 
structures that
correspond to dead worlds with no SASs there to behold them, 
for instance the 4+1-dimensional (atom-free) analog of our world, 
who cares whether they have PE or not?
In fact, as discussed in \cite{nihilo} and 
above in Section~\ref{EconomySec}, the totality of all worlds
is much less complex than such a specific world, 
and it is only in the subjective frog perspective of SASs
that seemingly complex structures such as trees and stellar 
constellations exist at all. Thus in this sense, not even 
the pines and the Big Dipper of our world would exist if 
neither we nor any other SASs
were here to perceive them.
The answer to Hawking's question 
``what is it that breathes fire into the equations and
makes a Universe for them to describe?'' \cite{HawkingBook}
would then be ``you, the SAS''.

Further attacking the distinction between physical and mathematical 
existence, one can speculate that future physicists will find 
the 1a TOE to be true {\it tautologically}\footnote{
Penrose has raised the following question \cite{Penrose}.
Suppose that a machine ``android" were built 
that simulated a human being 
so accurately that her friends could not tell the difference.
Would it be self-aware, {\ie}, subjectively feel like she does?
If the answer to this question is yes (in accordance with Leibniz'
``identity of indiscernibles" \cite{Tipler}), then as described below,
it might be untenable to make 
any distinction at all between physical existence and mathematical existence, 
so that our TOE would in fact be tautologically true.

Let us imagine a hypothetical Universe much larger than our own,
which contains a computer so powerful that it can simulate the
time-evolution of our entire Universe. By hypothesis, the
humans in this simulated world would perceive their world as being as 
real as we perceive ours, so by definition, the simulated universe would have PE.
Technical objections such as an infinite
quantity of information being required to store the data 
appear to be irrelevant to the philosophical point that we will make.
For instance, there is nothing about the physics we know today that
suggests that the Universe could not be replaced by a 
discrete and finite model that approximated it so closely that we, 
its inhabitants, could not tell the difference. 
That a vast amount of CPU-time would be needed is irrelevant, since 
that time bears no relation to the subjective time that the inhabitants of
the Universe would perceive. 
In fact, since we can choose to picture our Universe not as a 3D world
where things happen, but as a 4D world that 
merely is, there is no need for the computer to compute anything
at all --- it could simply store all the 4D data, and the 
``simulated" world would still have PE.
Clearly the way in which the data are stored should not matter, so
the amount of PE we attribute to the stored Universe should
be invariant under data compression. The physical laws provide a great
means of data compression, since they make it sufficient to store the initial 
data at some time together with the equations and an integration routine.
In fact, this should suffice even if the computer lacks
the CPU power and memory required to perform the decompression.
The initial data might be simple as well \cite{nihilo}, containing so
little algorithmic information that a single CD-ROM would
suffice to store it. After all, all that the needs to be stored 
is a description of the mathematical structure that is isomorphic 
to the simulated universe.
Now the ultimate question forces itself upon us: for this Universe 
to have PE, is the CD-ROM really needed at all? 
If this magic CD-ROM could be contained within the simulated Universe
itself, then it would ``recursively" support its own PE.
This would not involve any catch-22 ``hen-and-egg" problem
regarding whether the CD-ROM or the Universe existed first, 
since the Universe is a 4D structure which just is 
(``creation" is of course only a meaningful notion {\it within}
a spacetime).
In summary, a mathematical structure with SASs would have PE 
if it could be described purely formally (to a computer, say) 
--- and this is of course little else than having mathematical 
existence. 
Some closely related arguments can be found in \cite{Tipler}.
}.
This would be a natural extension of a famous analogy by Eddington
\cite{Eddington}:
\begin{itemize}
\item
One might say that wherever there is light, there are associated ripples
in the electromagnetic field. But the modern view is that 
light {\it is} the ripples.

\item 
One might say that wherever there is matter, there are associated ripples
in the metric known as curvature. But Eddington's view is that 
matter {\it is} the ripples.

\item
One might say that wherever there is PE, there is an associated mathematical
structure. But according to our TOE,
physical existence {\it is} the mathematical structure.

 
\end{itemize}

\subsection{And now what?}

If the TOE we have proposed is correct, then what are the implications?
As mentioned, answering the question of which mathematical structure we inhabit
requires ``business as usual" in terms of continuing theoretical and 
experimental research. However, there are also a number of implications 
for how we should focus our physics research, as summarized below.

\subsubsection{Don't mock e.g. string theorists}

In coffee rooms throughout the world, derogatory remarks are often heard
to the effect that certain physics theories 
(string theory, quantum groups, certain approaches to quantum gravity,
occasionally also grand unified and supersymmetric theories), 
are mere mathematical diversions, 
having nothing to do with physical reality.
According to our TOE, {\it all} such mathematical structures have physical 
existence if they are self-consistent, and therefore merit our study unless
it can be convincingly demonstrated that their properties preclude the
existence of SASs
\footnote{
Although it is often said that ``there are many perfectly good 
theories that simply turn out to be inconsistent with 
some experimental fact", this statement is certainly exaggerated.
The fact of the matter is that we to date have found {\it no} self-consistent
mathematical structure that can demonstrably describe both quantum and 
general relativistic phenomena. Classical physics was
certainly not ``a perfectly good theory", since it could not even
account for electromagnetism with sources in a
self-consistent way, and predicted that Hydrogen atoms
would collapse in a fraction of a second. In 1920, Herman Weyl remarked that
``the problem of matter is still shrouded in the deepest gloom"
\cite{Weyl}, and classical physics never became any better.
}.

\subsubsection{Compute probability distributions for everything}

One obvious first step toward more quantitative predictions from this TOE
is to explore the parameter space of the 
various continuous and discrete physical parameters 
in more detail, to map out the archipelago of
potential habitable islands.
For instance, by using a crude model for how the relevant nuclear spectra
depend on $\alpha$ and $\alpha_s$, one should attempt to map out 
the various islands allowing an unbroken reaction chain for 
heavy element production in stars, thereby refining \fig{alphaalphasFig}.
If the islands should turn out to cover only a tiny fraction of the
parameter space, it would become increasingly difficult to
believe in category 1b TOEs.
As mentioned, this type of calculation also offers 
a way to test and perhaps rule out
the TOE that we are proposing.

\subsubsection{Study the formal structure}

Since all mathematical structures are {\it a priori} 
given equal statistical weight, it is important
to study the purely formal nature of the mathematical 
models that we propose.
Would adding, changing or removing a few axioms 
produce an observationally equally viable model? For instance, 
if there is a large class of much more 
generic mathematical structures that would give identical predictions 
for low-energy physics but that would deviate at higher energies, 
then we would statistically expect to find 
such a deviation when our colliders
become able to probe these energy scales.
(Here we of course mean {\it mathematical} axioms, as in section~\ref{MEsec},
not informal ``axioms" regarding the correspondence between
the mathematics and our observations, such as 
``Axiom 3 of the Copenhagen interpretation of quantum mechanics".)

\subsubsection{Don't waste time on ``numerology"}

A popularly held belief is that 
the dimensionless parameters of nature will turn out
to be computable from first principles. For instance, 
Eddington spent years of his life developing theories where 
$\alpha$ was exactly $1/136$ and exactly $1/137$, respectively. 
If the above-mentioned analysis of heavy element production in stars were
to reveal that the archipelago of habitability covered only
a tiny fraction of the parameter space in \fig{alphaalphasFig},
then within the framework of our TOE, we would conclude that
it is highly unlikely that $\alpha$ is given by ``numerology", and 
concentrate our research efforts elsewhere.

\subsubsection{Shun classical prejudice}

Since all mathematical structures have PE in this theory, it of course
allows no sympathy whatsoever for subjective nostalgic bias towards structures
that resemble cozy classical concepts. For instance, criticism of
the Everett interpretation of quantum mechanics for being 
``too crazy" falls into this forbidden category.
Similarly, a model clearly cannot 
be criticized for involving ``unnecessarily large" ensembles.

\subsubsection{Don't neglect the frog perspective}

Considering how difficult it is to predict how a
mathematical structure will be perceived by a SAS,
a systematic study of this issue probably merits more 
attention than it is currently receiving. 
In the current era of specialization, where it is 
easy to get engrossed in mathematical technicalities, 
substantial efforts in phenomenology are needed
to make the connection with observation, going well beyond
simply computing cross sections and decay rates.

\subsection{Outlook}

Even if we are eventually able to figure out which mathematical structure 
we inhabit, which in the terminology of Weinberg \cite{Weinberg 1992} 
corresponds to discovering ``the final theory" (note that our usage
of `the word ``theory" is slightly different), our task as
physicists is far from over. In Weinberg's own words \cite{Weinberg 1992}:
{\it ``Wonderful phenomena, from turbulence to thought, will still 
need exploration whatever final theory is discovered.
The discovery of a final theory will not necessarily even help
very much in making progress in understanding these 
phenomena [...]. A final theory will be final in only one
sense --- it will bring to an end a certain 
sort of science, the ancient search for those principles that cannot
be explained in terms of deeper principles."}

If the TOE proposed in this paper is indeed correct, 
then the search for the ultimate principles has ended in
a slight anti-climax:
finding the TOE was easy, 
but working out its experimental implications is 
probably difficult enough to
keep physicists and mathematicians occupied 
for generations to come.

\acknowledgments 
The author wishes to thank Andreas Albrecht, Mark Alford,
John Barrow, Mats Boij, Martin Bucher,
Ted Bunn, Ang\'elica de Oliveira-Costa, 
Freeman Dyson, Piet Hut, Brit Katzen, Andrei Linde,
Dieter Maison, David Pittuck, Rainer Plaga, Bill Poirier, 
Georg Raffelt, Svend Rugh, Derek Storkey, 
Frank Tipler, John A. Wheeler, Frank Wilczek
and especially Harold Shapiro
for stimulating discussions on 
some of the above-mentioned topics.



\end{document}